\documentclass{IEEEtran}
\usepackage{cite}
\usepackage{amsmath,amssymb,amsfonts}
\usepackage{bbding}
\usepackage[caption=false,font=normalsize,labelfont=scriptsize,textfont=scriptsize]{subfig}
\usepackage{algorithmic}
\usepackage{graphicx}
\usepackage{textcomp}
\usepackage{multirow}
\usepackage{color}
\def\BibTeX{{\rm B\kern-.05em{\sc i\kern-.025em b}\kern-.08em
 T\kern-.1667em\lower.7ex\hbox{E}\kern-.125emX}}
\begin{document}
\title{Emerging Technologies in Intelligent Metasurfaces: Shaping the Future of Wireless Communications}
\author{Jiancheng An, \IEEEmembership{Member, IEEE}, M\'erouane Debbah, \IEEEmembership{Fellow, IEEE}, Tie Jun Cui, \IEEEmembership{Fellow, IEEE},\\Zhi Ning Chen, \IEEEmembership{Fellow, IEEE}, and Chau Yuen, \IEEEmembership{Fellow, IEEE} \\ \emph{\large{(Invited paper)}}
\thanks{This research is supported by National Research Foundation and Infocomm Media Development Authority under its Future Communications Research \& Development Programme (FCP-NTU-RG-2024-025), Science and Engineering Research Council of A*STAR (Agency for Science, Technology and Research) Singapore (Grant No. M22L1b0110), and Ministry of Education Singapore Tier2 (Award number T2EP50124-0032). T. J. Cui would like to acknowledge the National Natural Science Foundation of China (No. 62288101). \emph{(Corresponding author: Chau Yuen)}}
\thanks{J. An and C. Yuen are with the School of Electrical and Electronics Engineering, Nanyang Technological University (NTU), Singapore 639798 (e-mail: jiancheng.an@ntu.edu.sg, chau.yuen@ntu.edu.sg).}
\thanks{M. Debbah is with Khalifa University of Science and Technology, P O Box 127788, Abu Dhabi, UAE (e-mail: merouane.debbah@ku.ac.ae).}
\thanks{T. Cui is with the State Key Laboratory of Millimeter Wave, Southeast University, Nanjing 210096, China (e-mail: tjcui@seu.edu.cn).}
\thanks{Z. N. Chen is with the Department of Electrical and Computer Engineering, National University of Singapore (NUS), Singapore 117583 (e-mail: eleczn@nus.edu.sg).}}
\maketitle
\begin{abstract}
Intelligent metasurfaces have demonstrated great promise in revolutionizing wireless communications. One notable example is the two-dimensional (2D) programmable metasurface, which is also known as reconfigurable intelligent surfaces (RIS) to manipulate the wireless propagation environment to enhance network coverage. More recently, three-dimensional (3D) stacked intelligent metasurfaces (SIM) have been developed to substantially improve signal processing efficiency by directly processing analog electromagnetic signals in the wave domain. Another exciting breakthrough is the flexible intelligent metasurface (FIM), which possesses the ability to morph its 3D surface shape in response to dynamic wireless channels and thus achieve diversity gain. In this paper, we provide a comprehensive overview of these emerging intelligent metasurface technologies. We commence by examining recent experiments of RIS and exploring its applications from four perspectives. Furthermore, we delve into the fundamental principles underlying SIM, discussing relevant prototypes as well as their applications. Numerical results are also provided to illustrate the potential of SIM for analog signal processing. Finally, we review the state-of-the-art of FIM technology, discussing its impact on wireless communications and identifying the key challenges of integrating FIMs into wireless networks.
\end{abstract}

\section{Introduction}\label{S1}
\IEEEPARstart{P}{rogrammable} and intelligent metamaterials and metasurfaces have revolutionized the manipulations of electromagnetic (EM) fields and waves \cite{LSA_2014_Cui_Coding}. Typically, programmable metasurfaces are created by arranging an array of subwavelength-scale meta-atoms integrated with active components (e.g., PIN diodes, varactor diodes, micro-electromechanical systems) periodically on two-dimensional (2D) surfaces \cite{TAP_2022_Barbuto_Metasurfaces}. The digital coding representation and control of the coding pattern using the field programmable gate array (FGPA) have enabled the programmable metasurfaces to dynamically engineer the phase, amplitude, frequency, and polarization of EM waves in real time \cite{NC_2018_Zhang_Space, LSA_2019_Ma_Smart}. By manipulating the EM waves to an unprecedented degree, the metasurfaces exhibit extraordinary properties that do not exist in natural materials, such as negative refractive index, perfect absorption, and anomalous reflection/refraction patterns \cite{Science_2011_Yu_Light, Science_2014_Silva_Performing}. Over the past few decades, the metasurfaces have evolved from a theoretical concept to a fully-fledged field with a wide range of promising applications, including innovative waveguiding structures, absorbers, biomedical devices, and terahertz modulators and switches \cite{APM_2012_Holloway_An}.

\begin{figure}[!t]
\centerline{\includegraphics[width = 0.9 \columnwidth]{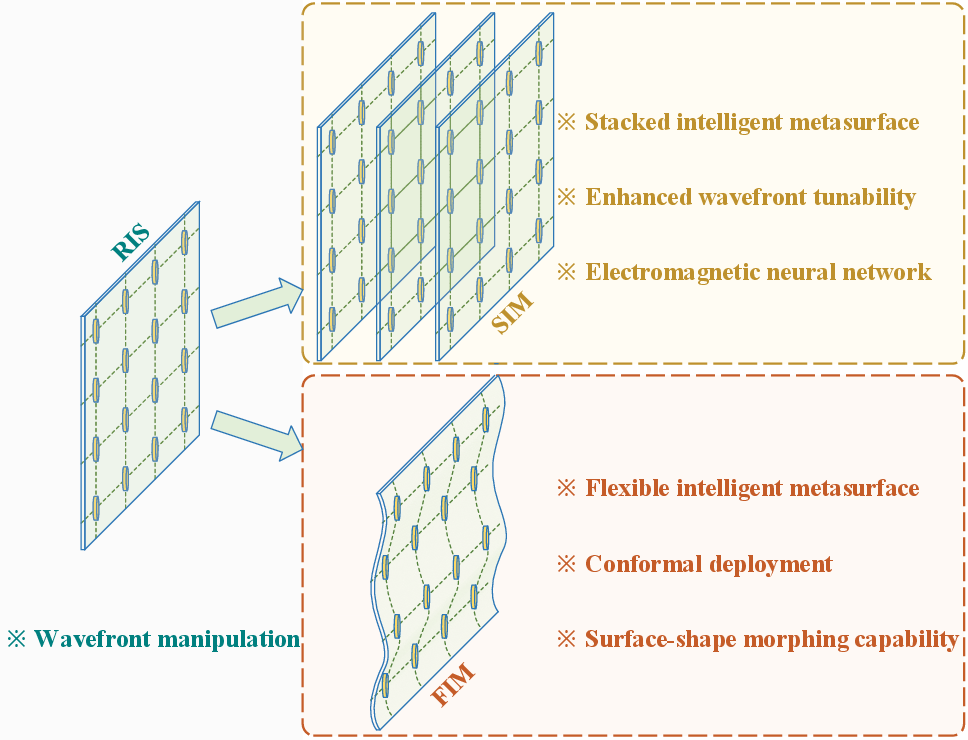}}
\caption{The development trend of intelligent metasurfaces: From 2D RIS to 3D SIM and FIM.}
\label{fig_1}
\end{figure}

\begin{table*}[]
\renewcommand\arraystretch{1.25}
\label{tab1}
\centering
\caption{Comparison of this work with existing surveys and tutorials on intelligent metasurfaces.}
\begin{tabular}{ll|c|c|c|c|c|c|c|c|c|c|c|c|l}
\hline
&\multicolumn{1}{l|}{Contents} & \makebox[0.03\textwidth][l]{\cite{TAP_2022_Barbuto_Metasurfaces}} & \makebox[0.03\textwidth][l]{\cite{JSAC_2020_Renzo_Smart}} & \makebox[0.03\textwidth][l]{\cite{CST_2021_Liu_Reconfigurable}} & \makebox[0.03\textwidth][l]{\cite{JSTSP_2022_Pan_An}} & \makebox[0.03\textwidth][l]{\cite{SPM_2022_Bjornson_Reconfigurable}} & \makebox[0.03\textwidth][l]{\cite{CST_2020_Gong_Toward}} & \makebox[0.03\textwidth][l]{\cite{CST_2022_Zheng_A}}& \makebox[0.03\textwidth][l]{\cite{VTM_2024_Basar_Reconfigurable}}& \makebox[0.03\textwidth][l]{\cite{WC_2024_An_Stacked}} &\makebox[0.03\textwidth][l]{\cite{arXiv_2024_Liu_Stacked}} & \makebox[0.03\textwidth][l]{\cite{JO_2020_Zang_Reconfigurable}} & \makebox[0.03\textwidth][l]{\cite{AN_2024_Zhou_Flexible}} & \makebox[0.075\textwidth][l]{This paper} \\ \hline\hline
\multicolumn{1}{l|}{\multirow{7}{*}{\textbf{RIS}}} & Hardware prototype & {\color{blue}{\CheckmarkBold}} & {\color{blue}{\CheckmarkBold}}& {\color{red}{\XSolidBrush}} & {\color{red}{\XSolidBrush}} & {\color{red}{\XSolidBrush}} & {\color{blue}{\CheckmarkBold}} & {\color{red}{\XSolidBrush}}& {\color{blue}{\CheckmarkBold}} & {\color{black}{---}} & {\color{black}{---}} & {\color{black}{---}} & {\color{black}{---}} & {\color{blue}{\CheckmarkBold}}\ \textbf{Sec. \ref{sec2_a}}\\ \cline{2-14} 
\multicolumn{1}{l|}{} & Field trials & {\color{blue}{\CheckmarkBold}} & {\color{red}{\XSolidBrush}}& {\color{red}{\XSolidBrush}} & {\color{red}{\XSolidBrush}} & {\color{red}{\XSolidBrush}} & {\color{blue}{\CheckmarkBold}} & {\color{red}{\XSolidBrush}} &{\color{blue}{\CheckmarkBold}} & {\color{black}{---}} & {\color{black}{---}} & {\color{black}{---}} & {\color{black}{---}} & {\color{blue}{\CheckmarkBold}}\ \textbf{Sec. \ref{sec2_a}} \\ \cline{2-14} 
\multicolumn{1}{l|}{} & Channel modeling & {\color{red}{\XSolidBrush}} & {\color{blue}{\CheckmarkBold}} & {\color{blue}{\CheckmarkBold}} & {\color{red}{\XSolidBrush}} & {\color{blue}{\CheckmarkBold}} & {\color{blue}{\CheckmarkBold}} & {\color{red}{\XSolidBrush}} & {\color{red}{\XSolidBrush}} & {\color{black}{---}} & {\color{black}{---}} & {\color{black}{---}} & {\color{black}{---}} & {\color{blue}{\CheckmarkBold}}\ \textbf{Sec. \ref{sec2_a}} \\ \cline{2-14}
\multicolumn{1}{l|}{} & Channel estimation & {\color{red}{\XSolidBrush}} & {\color{red}{\XSolidBrush}}& {\color{red}{\XSolidBrush}} & {\color{blue}{\CheckmarkBold}} & {\color{red}{\XSolidBrush}} & {\color{red}{\XSolidBrush}} & {\color{blue}{\CheckmarkBold}} & {\color{red}{\XSolidBrush}} & {\color{black}{---}} & {\color{black}{---}} & {\color{black}{---}} & {\color{black}{---}} & {\color{blue}{\CheckmarkBold}}\ \textbf{Sec. \ref{sec2_b}} \\ \cline{2-14}
\multicolumn{1}{l|}{} & Transmission design & {\color{red}{\XSolidBrush}}& {\color{red}{\XSolidBrush}} & {\color{blue}{\CheckmarkBold}} & {\color{blue}{\CheckmarkBold}} & {\color{blue}{\CheckmarkBold}} & {\color{blue}{\CheckmarkBold}} & {\color{blue}{\CheckmarkBold}} & {\color{red}{\XSolidBrush}}& {\color{black}{---}} & {\color{black}{---}} & {\color{black}{---}} & {\color{black}{---}} & {\color{blue}{\CheckmarkBold}}\ \textbf{Sec. \ref{sec2_b}} \\ \cline{2-14} 
\multicolumn{1}{l|}{} & Radio localization & {\color{red}{\XSolidBrush}} & {\color{red}{\XSolidBrush}}& {\color{red}{\XSolidBrush}} & {\color{blue}{\CheckmarkBold}} & {\color{blue}{\CheckmarkBold}} & {\color{red}{\XSolidBrush}}& {\color{red}{\XSolidBrush}} & {\color{red}{\XSolidBrush}} & {\color{black}{---}} & {\color{black}{---}} & {\color{black}{---}} & {\color{black}{---}} & {\color{red}{\XSolidBrush}}\\ \cline{2-14}
\multicolumn{1}{l|}{} & Resource management & {\color{red}{\XSolidBrush}}& {\color{red}{\XSolidBrush}} & {\color{blue}{\CheckmarkBold}} & {\color{red}{\XSolidBrush}} & {\color{red}{\XSolidBrush}} & {\color{red}{\XSolidBrush}} & {\color{red}{\XSolidBrush}} & {\color{red}{\XSolidBrush}} & {\color{black}{---}} & {\color{black}{---}} & {\color{black}{---}} & {\color{black}{---}} & {\color{red}{\XSolidBrush}}\\ \hline\hline
\multicolumn{1}{l|}{\multirow{4}{*}{\textbf{SIM}}} & Operating principle& {\color{black}{---}} & {\color{black}{---}} & {\color{black}{---}} & {\color{black}{---}} & {\color{black}{---}}& {\color{black}{---}} & {\color{black}{---}} & {\color{black}{---}} & {\color{blue}{\CheckmarkBold}} & {\color{blue}{\CheckmarkBold}} & {\color{black}{---}}& {\color{black}{---}} & {\color{blue}{\CheckmarkBold}}\ \textbf{Sec. \ref{sec3_a}} \\ \cline{2-14} 
\multicolumn{1}{l|}{} & Hardware prototype & {\color{black}{---}} & {\color{black}{---}} & {\color{black}{---}} & {\color{black}{---}} & {\color{black}{---}}& {\color{black}{---}} & {\color{black}{---}} & {\color{black}{---}} & {\color{red}{\XSolidBrush}} & {\color{blue}{\CheckmarkBold}} & {\color{black}{---}} & {\color{black}{---}} & {\color{blue}{\CheckmarkBold}}\ \textbf{Sec. \ref{sec3_b}} \\ \cline{2-14} 
\multicolumn{1}{l|}{} & Promising application & {\color{black}{---}} & {\color{black}{---}} & {\color{black}{---}} & {\color{black}{---}} & {\color{black}{---}}& {\color{black}{---}} & {\color{black}{---}} & {\color{black}{---}} & {\color{red}{\XSolidBrush}} & {\color{red}{\XSolidBrush}} & {\color{black}{---}} & {\color{black}{---}} & {\color{blue}{\CheckmarkBold}}\ \textbf{Sec. \ref{sec3_c}} \\ \cline{2-14} 
\multicolumn{1}{l|}{} & Performance evaluation & {\color{black}{---}} & {\color{black}{---}} & {\color{black}{---}} & {\color{black}{---}} & {\color{black}{---}}& {\color{black}{---}} & {\color{black}{---}} & {\color{black}{---}} & {\color{blue}{\CheckmarkBold}} & {\color{red}{\XSolidBrush}} & {\color{black}{---}} & {\color{black}{---}} & {\color{blue}{\CheckmarkBold}}\ \textbf{Sec. \ref{sec3_e}} \\ \hline\hline
\multicolumn{1}{l|}{\multirow{3}{*}{\textbf{FIM}}} & Hardware prototype &{\color{black}{---}} & {\color{black}{---}} & {\color{black}{---}} & {\color{black}{---}} & {\color{black}{---}}& {\color{black}{---}} & {\color{black}{---}} & {\color{black}{---}} & {\color{black}{---}} & {\color{black}{---}} & {\color{blue}{\CheckmarkBold}} & {\color{blue}{\CheckmarkBold}}& {\color{blue}{\CheckmarkBold}}\ \textbf{Sec. \ref{sec4_a}} \\ \cline{2-14} 
\multicolumn{1}{l|}{} & Performance evaluation & {\color{black}{---}} & {\color{black}{---}} & {\color{black}{---}} & {\color{black}{---}} & {\color{black}{---}}& {\color{black}{---}} & {\color{black}{---}} & {\color{black}{---}} & {\color{black}{---}} & {\color{black}{---}} & {\color{red}{\XSolidBrush}} & {\color{red}{\XSolidBrush}} & {\color{blue}{\CheckmarkBold}}\ \textbf{Sec. \ref{sec4_c}} \\ \cline{2-14} 
\multicolumn{1}{l|}{} & Open challenge & {\color{black}{---}} & {\color{black}{---}} & {\color{black}{---}} & {\color{black}{---}} & {\color{black}{---}}& {\color{black}{---}} & {\color{black}{---}} & {\color{black}{---}} & {\color{black}{---}} & {\color{black}{---}} & {\color{red}{\XSolidBrush}} & {\color{red}{\XSolidBrush}} & {\color{blue}{\CheckmarkBold}}\ \textbf{Sec. \ref{sec4_d}} \\ \hline
\end{tabular}
\end{table*}
The recent innovative application of programmable metasurfaces in wireless communications has sparked considerable interest \cite{JSAC_2020_Renzo_Smart, WC_2024_An_Codebook, TWC_2019_Huang_Reconfigurable, TCOM_2021_Wu_Intelligent}. Traditionally, the wireless propagation environment has been viewed as uncontrollable, impacting the efficiency and reliability of wireless transmission. However, programmable metasurfaces, commonly referred to as reconfigurable intelligent surfaces (RIS) in this context, have refuted this perspective. By intelligently coordinating the amplitude and phase shifts of incident EM waves through a large number of low-cost, quasi-passive reflecting elements, RIS can manipulate the radio propagation environment on demand \cite{TCOM_2021_Wu_Intelligent}. Compared to conventional relays, RIS offers an attractive advantage in terms of lower power consumption, as it no longer requires expensive and power-hungry radio-frequency (RF) components for signal transmission \cite{TWC_2019_Huang_Reconfigurable}. Furthermore, RIS is typically lightweight and compact, making it easy to integrate with various environment structures, such as buildings, ceilings, light poles, and road signs \cite{CST_2021_Liu_Reconfigurable}. In past years, substantial research efforts have demonstrated the unprecedented performance gains of RIS for enhancing the capacity and coverage of wireless networks in a cost-effective and energy-efficient manner \cite{TCOM_2021_Wu_Intelligent, TWC_2019_Huang_Reconfigurable, TCOM_2022_An_Low}.

Recently, three-dimensional (3D) \textbf{stacked intelligent metasurfaces (SIM)} have been developed to realize analog signal processing in the wave domain \cite{WC_2024_An_Stacked}. As shown in Fig. \ref{fig_1}, a SIM is physically fabricated by closely packing multiple layers of programmable metasurfaces, resulting in an architecture that bears some conceptual resemblance to an artificial neural network (ANN) \cite{Science_2018_Lin_All}. The electronically programmable meta-atoms within a SIM function similarly to artificial neurons in an ANN, with their complex-valued transmission coefficients serving as trainable network parameters \cite{NE_2022_Liu_A}. By properly configuring the transmission coefficients of the meta-atoms on each metasurface, a SIM becomes capable of automatically accomplishing various advanced computation and signal processing tasks as EM waves propagate through the layered structure at the speed of light \cite{JSAC_2023_An_Stacked, JSAC_2024_An_Two}. Since SIMs directly process the information-carrying EM waves in free space, they no longer require any digital storage, transmission, pre-processing of information, as well as external computing power. Recent prototypes have demonstrated SIMs with impressive capabilities for applications such as MIMO beamforming \cite{JSAC_2023_An_Stacked}, radar sensing \cite{JSAC_2024_An_Two}, and image classification \cite{LSA_2022_Luo_Metasurface}.

In addition, \textbf{flexible intelligent metasurfaces (FIM)} have been created by depositing an array of dielectric inclusions onto a soft, conformal substrate such as polydimethylsiloxane (PDMS) \cite{JO_2020_Zang_Reconfigurable, AN_2024_Zhou_Flexible}. More recently, a shape-morphable FIM has been developed using a mesh of tiny metallic filaments that can be precisely controlled by reprogrammable Lorentz forces generated by electrical currents passing through a static magnetic field \cite{Nature_2022_Bai_A}. This type of FIM can morph its 3D surface shape into a wide variety of target configurations within milliseconds. The application of FIMs to wireless networks has shown significant potential, especially in terms of their adaptability to dynamic wireless environments \cite{TWC_2024_An_Flexible, TCOM_2025_An_Flexible}. Specifically, conventional 2D rigid antenna arrays with fixed element positions may lead to poor channel quality, especially when deep fading occurs \cite{Book_2005_Tse_Fundamentals}. However, by strategically adjusting the physical position of each antenna element, FIMs can ensure that multiple signal copies from different paths add constructively, thus improving the received signal quality through beneficial 3D surface-shape morphing \cite{TWC_2024_An_Flexible}. This is particularly beneficial for millimeter-wave and terahertz communications, where the channel coherence distance is relatively small.

In this article, we provide a comprehensive overview of the state-of-the-art of intelligent metasurfaces in wireless communications. First, we review the latest research breakthroughs in experimental testing and modeling of RIS. Several key developments are discussed from four different perspectives. In the second part, we delve into the fundamental operating principles for SIM and systematically survey existing prototype designs. Moreover, we identify promising application scenarios for integrating SIM into future wireless networks and provide numerical results to demonstrate their advantages. Finally, we discuss the emerging FIM technology and its potential impacts on wireless communications. We also provide insights into ongoing challenges and outline major research directions that warrant further investigation. While several recent survey papers \cite{AP_2024_Bilotti_Reconfigurable, JSAC_2020_Renzo_Smart, CST_2021_Liu_Reconfigurable, VTM_2024_Basar_Reconfigurable, JSTSP_2022_Pan_An, SPM_2022_Bjornson_Reconfigurable, CST_2020_Gong_Toward, CST_2022_Zheng_A} have examined the same topic, they have primarily focused on 2D RIS. In contrast, this survey places a greater emphasis on 3D SIM and FIM. More explicitly, we summarize the unique contributions of this paper compared to existing works in Table \ref{tab1}, where the structure of the paper is also outlined.
\begin{figure*}[!t]
\centerline{\includegraphics[width = 1.9 \columnwidth]{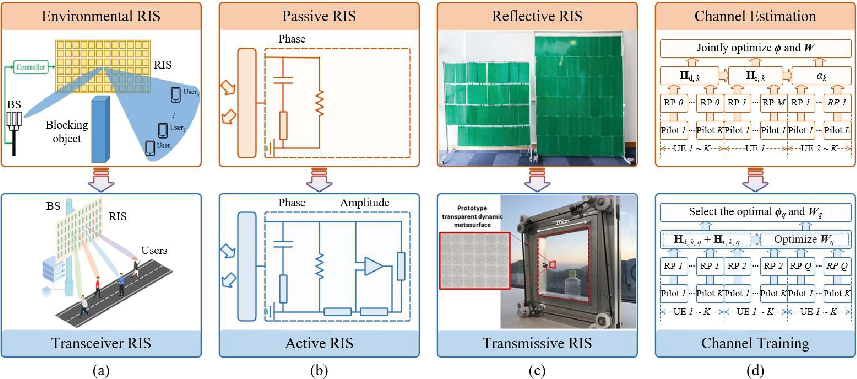}}
\caption{Four development trends of RIS technology: (a) from environmental RIS to transceiver RIS; (b) from passive RIS to active RIS; (c) from reflective RIS to transmissive RIS; (d) from channel estimation to channel training.}\label{fig_2}
\end{figure*}

\section{Reconfigurable Intelligent Surface (RIS)}
In this section, we first review the recent progress in modeling and experimental field tests of RIS (Section \ref{sec2_a}). Then we discuss the applications of RISs in wireless communications from four distinct perspectives (Section \ref{sec2_b}).
\subsection{Experiment and Modeling}\label{sec2_a}
\subsubsection{\bf{Prototype and Field Trials}}
Over the past years, various prototypes of RIS have been developed and experimentally tested to showcase their potential in enhancing wireless communications. \emph{Liu et al.} \cite{PRA_2019_Liu_Intelligent} designed a reflection-type RIS operating in the microwave frequency range, which is capable of achieving local control over the complex surface impedance for tunable perfect absorption and anomalous reflection. In \cite{Access_2020_Dai_Reconfigurable}, \emph{Dai et al.} developed a high-gain yet cost-effective RIS with $256$ elements that integrated phase shifting and radiation functions using PIN diodes, allowing for $2$-bit phase shift tuning. Testing at $28.5$ GHz revealed an impressive antenna gain of $19.1$ dBi. In \cite{TCOM_2021_Pei_RIS}, a RIS prototype with $1,100$ individually controllable elements operating at $5.8$ GHz was presented. In an indoor test scenario where the transmitter and receiver are separated by a $30$ cm concrete wall, the RIS provides a power gain of $26$ dB compared to using a copper surface alone. \emph{Huang et al.} \cite{TAP_2021_Huang_Active} designed an active frequency selective surface (AFSS) consisting of a periodic metallic structure layer and a graphene capacitor layer, which enables EM switching between transmission and absorption modes within the $6\sim7$ GHz band \cite{TAP_2021_Huang_Active}.

Additionally, the authors of \cite{OJCOMS_2022_Trichopoulos_Design} designed and fabricated a RIS with $160$ elements operating at $5.8$ GHz. In realistic outdoor communication scenarios with a blockage between a base station (BS) and a grid of mobile users over a $35$ m propagation path, the RIS demonstrates an average signal-to-noise ratio (SNR) improvement of more than $6$ dB. Recently, \emph{Liang et al.} \cite{TAP_2022_Liang_An} designed a $3$-bit meta-atom that is insensitive to incident angles. A RIS fabricated using this angle-insensitive meta-atom maintains stable phase and amplitude responses over a wide incidence angle range of $\pm\, 60^{\circ}$, verifying the channel reciprocity in RIS-aided wireless networks \cite{TAP_2022_Liang_An}. In \cite{OJCOMS_2024_Zhao_2}, \emph{Zhao et al.} developed a RIS architecture that enables full-array space-time wavefront manipulation. Notably, each meta-atom offers $2$-bit phase-shifting capabilities and exhibits an insertion loss of less than $0.6$ dB across a $200$ MHz bandwidth. Moreover, these experimental results have demonstrated the potential of RIS to significantly reduce power consumption in real-world communication systems. Inspired by digital electronics, \emph{Yang et al.} \cite{TAP_2024_Yang_High} synthesized a metasurface with arbitrary desired phases as effortlessly as Boolean algebra. They demonstrated the capability of the designed metasurface for beamforming, double beam generation, and dynamic vortex beam creation.

\subsubsection{\bf{Modeling}}
Developing mathematically tractable yet reasonably accurate EM models is crucial for theoretically analyzing and optimizing RIS-aided communication systems. To this end, the authors of \cite{TCOM_2021_Najafi_Physics} developed a physics-based model by partitioning the RIS elements into several \emph{tiles}. They modeled each tile as an anomalous reflector and then derived its impact by solving the resulting integral equations for the electric and magnetic vector fields. Moreover, \emph{Bj$\ddot{o}$rnson} and \emph{Sanguinetti} \cite{WCL_2021_Bjornson_Rayleigh} provided a physically feasible Rayleigh fading model that accurately characterizes the correlation between channels associated with different RIS elements, while \emph{Di Renzo et al.} \cite{Proc_2022_Renzo_Communication} developed a reflection model based on inhomogeneous sheets theory with varying surface impedance. Additionally, \emph{Tang et al.} \cite{TCOM_2022_Tang_Path} developed a free-space path loss model for RISs, which takes into account the directivity of the transceiver antenna and reflecting element through an angle-dependent loss factor. By leveraging rigorous scattering parameter network analysis, \emph{Shen et al.} \cite{TWC_2022_Shen_Modeling} proposed two new RIS architectures based on group- and fully-connected impedance networks, which are shown to increase received signal power by more than $60\%$ compared to single-connected impedance networks.

\subsection{Development Trend}\label{sec2_b}
In this subsection, we will discuss the development trend of RIS technology from four perspectives: i) deployment location, ii) signal enhancement capability, iii) coverage area, and iv) practical protocol.
\subsubsection{\bf{From Environmental RIS to Transceiver RIS}}
One of the main applications of RISs is to deploy them in propagation environments to assist in wireless communications \cite{TCOM_2021_Wu_Intelligent, TCOM_2022_An_Low}. In recent years, substantial research efforts have demonstrated the potential of utilizing RIS to enhance the energy efficiency of cellular networks \cite{TWC_2019_Huang_Reconfigurable}, increase the capacity of orthogonal frequency division multiplexing systems \cite{TVT_2024_An_Adjustable}, bolster physical layer security \cite{JSAC_2020_Yu_Robust}, and reduce latency in mobile edge computing \cite{JSAC_2020_Bai_Latency}, to name just a few.

In addition, RIS can be integrated with transceivers to make use of the entire surface for both transmission and reception, as shown in Fig. \ref{fig_2}(a), enabling holographic MIMO (HMIMO) communications \cite{JSAC_2020_Pizzo_Spatially, TSP_2018_Hu_Beyond}. This integration brings two distinct benefits:
\begin{itemize}
 \item \emph{Near-Field Communications:} When a large intelligent surface is utilized, users are typically within its radiating near-field region. This enables the exploitation of full spatial degrees of freedom (DoF), even in strong line-of-sight (LoS) channel conditions \cite{WC_2024_An_Near}.
 \item \emph{Continuous Aperture:} By integrating a near-infinite number of radiating and sensing elements into a finite space, a spatially continuous surface is formed. This allows for leveraging the spatial diversity gain across the whole continuous surface\footnote{Motivated by this philosophy, movable and fluid antennas have been utilized in wireless networks to attain spatial diversity \cite{TWC_2021_Wong_Fluid, TWC_2024_Zhu_Modeling}.} \cite{CL_2023_An_A}.
\end{itemize}

To characterize the fundamental performance limits of HMIMO communications, \emph{Hu et al.} \cite{TSP_2018_Hu_Beyond} considered a LoS propagation environment and demonstrated that the maximum spatial DoF is $\pi/\lambda^{2}$ per square meter for 2D terminal deployments. Based on the scalar Helmholtz equation, \emph{Pizzo et al.} \cite{JSAC_2020_Pizzo_Spatially} modeled the 3D small-scale fading in HMIMO communications as a Fourier plane-wave spectral representation. Furthermore, \emph{Jung et al.} \cite{TWC_2020_Jung_Performance} showed the \emph{channel hardening} phenomenon in multiuser HMIMO systems. An asymptotic analysis of the uplink data rate indicates that the impact of hardware impairments, channel estimation errors, and noise becomes negligible as the number of antennas and devices increases. Additionally, \emph{Deng et al.} \cite{JSAC_2022_Deng_HDMA} proposed a holographic-pattern division multiple access strategy that maps the intended signals to a superposed holographic pattern, while \emph{Wei et al.} \cite{TWC_2023_Wei_Tri} utilized triple polarization patch antennas at transceivers to enhance the capacity and diversity of multiuser HMIMO communication systems. More recently, \emph{Dardari} \cite{arXiv_2024_Dardari_Dynamic} introduced the concept of a dynamic scattering array (DSA) to enable efficient HMIMO communications, shifting signal processing from the digital domain to the EM domain. Please refer to \cite{CL_2023_An_A, CST_2024_Gong_Holographic} for a comprehensive survey on the latest advances in HMIMO communications.

\subsubsection{\bf{From Passive RIS to Active RIS}}
Due to the two-hop propagation -- from the transmitter to the RIS and then from the RIS to the receiver -- the overall path loss in RIS-aided communication systems is severe. To compensate for the multiplicative fading attenuation, a large number of passive reflecting elements are often deployed, resulting in a large surface area and increased circuit power consumption. To address this issue, \emph{Long et al.} \cite{TWC_2021_Long_Active} proposed a new type of active RIS that can not only reflect signals with adjustable phase shifts but also amplify incoming signals. They also developed a method for optimizing the reflection coefficients to minimize RIS-related noise, while maximizing the received signal power. Moreover, \emph{Zhang et al.} \cite{TCOM_2023_Zhang_Active} designed an active RIS by integrating a small amplifier into each element, as depicted in Fig. \ref{fig_2}(b). They developed and experimentally verified a signal model to characterize the signal amplification. Based on the tested model, they demonstrated the potential of active RISs to double the sum rate for multiuser MISO communications.

\subsubsection{\bf{From Reflective RIS to Transmissive RIS}}
Most studies on RIS have concentrated on reflective metasurfaces that reflect incoming signals toward receivers on the same side as the transmitter. However, this limits the service coverage to only half of the surrounding space of RIS. To overcome this limitation, simultaneously transmitting and reflecting (STAR) RISs were conceived to refract and reflect wireless signals into the entire space, as shown in Fig. \ref{fig_2}(c). In \cite{TWC_2022_Mu_Simultaneously}, \emph{Mu et al.} proposed three practical protocols for operating STAR-RISs: energy splitting, mode switching, and time switching. They also explored the downlink of a STAR-RIS-aided multiuser MISO communication system, where the transmit beamforming and the passive transmission/reflection beamforming at the STAR-RIS were jointly optimized to minimize power consumption for both unicast and multicast transmissions. Similarly, \emph{Zhang et al.} \cite{CM_2022_Zhang_Intelligent, TWC_2022_Zhang_Intelligent} introduced the concept of intelligent omni-surface (IOS) and presented a relevant prototype. A hybrid downlink beamforming scheme was proposed to maximize the sum rate of multiple users on both sides of the IOS. Numerical results have demonstrated that STAR-RISs and IOSs can significantly extend the service coverage of a BS compared to conventional transmitting/reflecting-only RISs.

\subsubsection{\bf{From Channel Estimation to Channel Training}}
In RIS-aided communication systems, accurate channel state information (CSI) acquisition and feedback with moderate overhead presents a major challenge. This is because passive RIS elements are unable to transmit or receive pilot signals. To address this issue, advanced channel estimation techniques have been developed to probe the cascaded transmitter-RIS-receiver channels. Specifically, \emph{You et al.} \cite{JSAC_2020_You_Channel} proposed a hierarchical protocol design to progressively estimate the reflection channels over multiple time blocks by exploiting grouping strategies and considering the discrete phase shift constraint. Moreover, \emph{Zhi et al.} \cite{TIT_2023_Zhi_Two} developed a two-timescale transmission scheme for RIS-aided massive MIMO systems. The transmit beamforming is adapted to the rapidly changing instantaneous CSI, while the passive beamforming is adapted to the slowly changing statistical CSI. Additionally, the authors of \cite{TVT_2023_Xu_Channel} and \cite{TCCN_2022_Xu_Time} investigated estimation and prediction for rapidly time-varying channels, respectively, while \cite{TSP_2020_Zhou_A} studied robust passive beamforming design based on imperfect CSI.

However, the pilot overhead in the aforementioned channel estimation schemes still increases with the number of RIS elements. To reduce pilot signaling overhead and implementation complexity, the authors of \cite{TCOM_2022_An_Low, TGCN_2022_An_Joint} proposed a new channel training protocol, which is also known as \emph{codebook} solution. As shown in Fig \ref{fig_2}(d), in each training block, the RIS is configured based on a predesigned codebook. The composite channel is then estimated, based on which the transmit beamformer is designed with low complexity. The codeword resulting in optimal performance is used to assist data transmission \cite{WC_2024_An_Codebook}. Substantial research efforts have shown that the codebook protocol is more robust against channel estimation errors \cite{TCOM_2022_An_Low, TGCN_2022_An_Joint}. Recently, \emph{Ren et al.} \cite{TWC_2023_Ren_Configuring} proposed a blind beamforming strategy for RIS that extracts statistical features from random samples of the received signal power. Additionally, \emph{Yu et al.} \cite{TCOM_2024_Yu_Environment} proposed an environment-aware codebook protocol, which generates a reflection pattern codebook based on statistical CSI, thus achieving better performance than conventional environment-agnostic codebooks.

\begin{figure}[!t]
\centerline{\includegraphics[width = 0.9 \columnwidth]{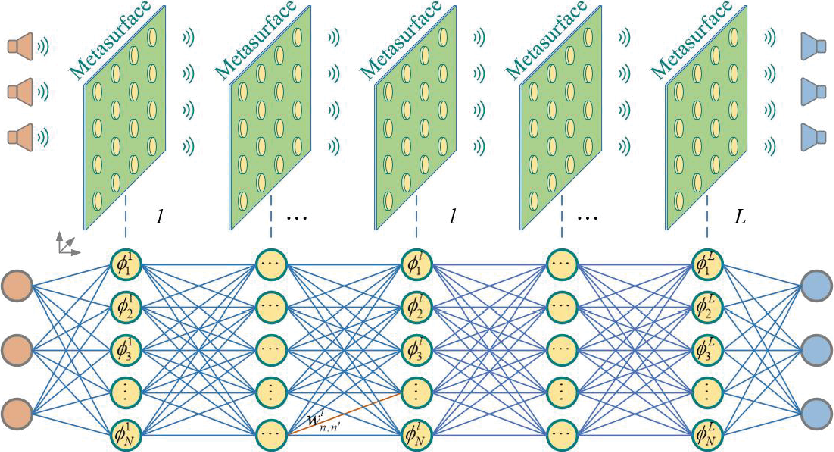}}
\caption{Schematic of a SIM and its equivalent neural network structure. The forward propagation of EM waves in the SIM is equivalent to a multi-layer perceptron.}
\label{fig_3}
\end{figure}
\section{Stacked Intelligent Metasurface (SIM)}
In this section, we shift our focus to 3D SIM technology, which provides a new paradigm to perform signal processing. Specifically, we first elaborate on the fundamental operating principle of SIM in Section \ref{sec3_a}, and then discuss some existing prototypes in Section \ref{sec3_b}. Furthermore, Section \ref{sec3_c} reviews the functionalities of SIM, while Section \ref{sec3_d} provides a comprehensive survey of the state-of-the-art in applying SIM technology to wireless networks. In Section \ref{sec3_e}, we provide case studies to show the capabilities of SIM and shed light on the future directions in Section \ref{sec3_f}.

\subsection{Operating Principle}\label{sec3_a}
As illustrated in Fig. \ref{fig_3}, a SIM is constructed by stacking multiple layers of programmable metasurfaces. Each layer contains a large number of low-cost, passive meta-atoms that can manipulate EM waves. Essentially, the architecture of a SIM forms an EM neural network (EMNN), which is conceptually similar to an ANN \cite{Science_2018_Lin_All, NE_2022_Liu_A}. Specifically, each intelligent metasurface can be viewed as a hidden layer, where the meta-atoms behave similarly to artificial neurons in a conventional ANN. By adequately training and configuring the transmission coefficients of the meta-atoms using error backpropagation, a SIM is capable of performing desired functions through the interactions of spatial EM waves with the metasurface layers \cite{WC_2024_An_Stacked, arXiv_2024_Liu_Stacked}. In recent years, several advanced SIM architectures have been reported to demonstrate impressive capabilities for various emerging applications, such as hologram generation \cite{TAP_2024_Jia_High}, image classification \cite{LSA_2022_Luo_Metasurface}, and matrix operations \cite{JSAC_2023_An_Stacked, JSAC_2024_An_Two}.

According to the Huygens–Fresnel principle \cite{Science_2018_Lin_All}, each meta-atom on a metasurface layer acts as a secondary source, producing a spherical wave that illuminates all the meta-atoms on the next layer. The field response impinging on each layer is the weighted sum of the diffracted fields from the meta-atoms on the previous layer, with the weight coefficient determined by the product of the complex-valued transmission coefficients and corresponding propagation coefficients \cite{JSAC_2023_An_Stacked, JSAC_2024_An_Two}. Based on the Rayleigh–Sommerfeld diffraction integral, the propagation coefficient from the $n'$-th meta-atom on the $\left ( l-1 \right )$-th metasurface layer to the $n$-th meta-atom on the $l$-th metasurface layer can be expressed by\footnote{In practical systems, the inter-layer propagation coefficients may be inconsistent with \eqref{eq1} due to inevitable manufacturing errors and misalignment among multiple metasurfaces, which can be calibrated by leveraging the backward propagation algorithm before configuring the SIM \cite{JSAC_2023_An_Stacked}.}
\begin{align}\label{eq1}
 w_{n,{n}'}^{l}=\frac{A\cos\chi_{n,{n}'}^{l} }{r_{n,{n}'}^{l}}\left ( \frac{1}{2\pi r_{n,{n}'}^{l}}-j\frac{1}{\lambda } \right )e^{j2\pi r_{n,{n}'}^{l}/\lambda },
\end{align}
where $A$ is the area of each meta-atom, $r_{n,{n}'}^{l}$ denotes the corresponding propagation distance, while $\chi_{n,{n}'}^{l}$ represents the angle between the propagation direction and the normal direction of the $\left ( l-1 \right )$-th metasurface layer. By repeating this propagation process until the last metasurface layer, the transfer function $\mathbf{S}$ of the SIM can be formulated as\footnote{In a practical SIM device, reflective coatings are often applied to reduce inter-layer reflections inside it, thus ensuring the forward propagation of EM signals.}
\begin{align}\label{eq2}
 \mathbf{S} = \mathbf{\Phi} ^{L}\mathbf{W}^{L}\cdots \mathbf{\Phi} ^{2}\mathbf{W}^{2}\mathbf{\Phi} ^{1}\in \mathbb{C}^{N\times N},
\end{align}
where $N$ is the number of meta-atoms on each metasurface layer, $\mathbf{W}^{l}\in \mathbb{C}^{N\times N}$ denotes the propagation coefficient matrix between the $\left ( l-1 \right )$-th metasurface layer and the $l$-th metasurface layer, and $\mathbf{\Phi}^{l}\in \mathbb{C}^{N\times N}$ represents the transmission coefficient matrix of the $l$-th metasurface layer.

 \begin{figure*}[!t]
\centerline{\includegraphics[width = 1.9\columnwidth]{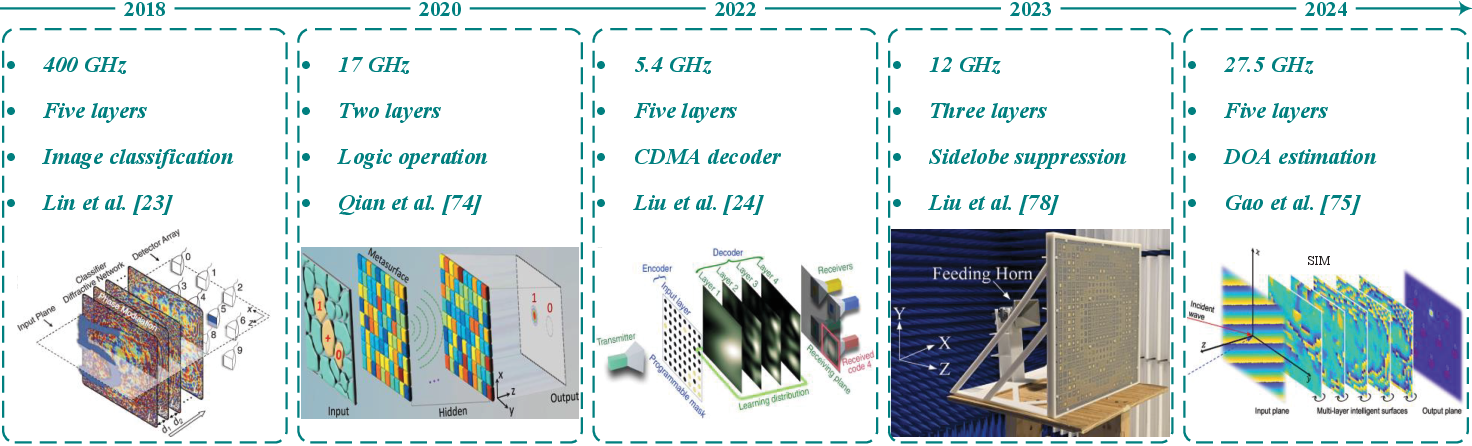}}
\caption{Illustration of several existing SIM prototypes.}
\label{fig_4}
\end{figure*}
Since the matrix operations in \eqref{eq2} occur automatically during the wave propagation through free space, SIMs can execute complex signal processing and computing tasks without requiring any digital storage, transmission, or preprocessing of information \cite{NE_2022_Liu_A}. Additionally, SIM can perform parallel calculations for all input EM signals, demonstrating its excellent scalability \cite{WC_2024_An_Stacked}.

\subsection{Hardware Prototypes}\label{sec3_b}
As shown in Fig. \ref{fig_4}, several SIM prototypes have been developed over the past years, turning the abstract concept into a tangible reality. In \cite{Science_2018_Lin_All}, \emph{Lin et al.} designed a passive SIM using 3D printed diffractive layers capable of classifying images of handwritten digits and fashion products at terahertz frequencies. Building on this, \emph{Qian et al.} \cite{LSA_2020_Qian_Performing} created a compact SIM that can perform logic operations (NOT, OR, and AND) at microwave frequencies. The plane wave is initially spatially encoded using a patterned mask according to input logic states and then passes through a series of hidden metasurface layers. The well-designed SIM scatters the encoded EM waves into one of two designated areas at the output layer, indicating the output logic state.

Furthermore, \emph{Liu et al.} \cite{NE_2022_Liu_A} developed a reconfigurable SIM using programmable metasurfaces that can dynamically adjust its EM behavior in real time. Specifically, each meta-atom (neuron) is connected to two amplifier chips, providing a dynamic modulation range of $35$ dB. A five-layer SIM prototype was fabricated to demonstrate its capability for image recognition and multi-beam focusing \cite{NE_2022_Liu_A}. To enhance the parallelism capability of SIM, \emph{Luo et al.} \cite{LSA_2022_Luo_Metasurface} developed a SIM device that can simultaneously recognize digital and fashionable items by leveraging polarization multiplexing in the visible spectrum. By integrating the SIM with a standard imaging sensor, a chip-scale architecture was created for energy-efficient and ultra-fast visual information processing. Recently, \emph{Jia et al.} \cite{TAP_2024_Jia_High} fabricated a two-layer SIM by using tunable transmissive metasurfaces with opposite phases, which can achieve both information transmission and holographic image generation using multiple channels. Additionally, \emph{Gao et al.} \cite{LSA_2024_Gao_Super} designed a SIM for estimating the direction-of-arrival (DOA) of multiple radio sources over a $5$ GHz bandwidth. Spatial-temporal multiplexing was utilized to achieve high-resolution DOA estimation over a wide field of view. In \cite{arXiv_2024_Wang_Multi}, a SIM comprised of three transmissive metasurface layers was built for enabling integrated sensing and communication (ISAC) tasks at $5.8$ GHz. Moreover, \emph{Liu et al.} \cite{TAP_2022_Liu_Prior, TAP_2023_Liu_Full} synthesized a triple-layer meta-atom that enables precise phase shift control across a range of $330^{\circ}$, with a signal loss of less than $-1$ dB. This breakthrough has facilitated the design of metalens that can effectively suppress sidelobes. In parallel, \emph{Lv et al.} \cite{Nano_2024_Lv_Meta} have explored the potential of SIM technology to produce a nearly rectangular filtering response.

\begin{figure*}[!t]
\centerline{\includegraphics[width = 1.9\columnwidth]{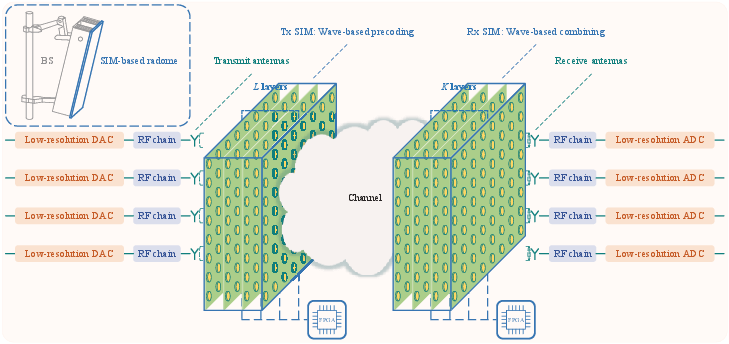}}
\caption{A SIM-aided point-to-point MIMO system, where a pair of SIMs are deployed at the transmitter and receiver to implement MIMO precoding and combining automatically as the EM waves propagate through them.}\label{fig_5}
\end{figure*}

\subsection{Functionalities of SIM in Wireless Communications}\label{sec3_c}
Leveraging SIM to achieve wave-domain signal processing has great potential to significantly reduce hardware costs and energy consumption of communication systems. In this subsection, we first review recent research efforts of utilizing SIM for performing beamforming-related matrix operations in the wave domain. Then we discuss the contributions related to channel estimation and machine learning (ML) based design.
\subsubsection{\bf{MIMO Beamforming}}
Recent studies have made considerable progress in utilizing SIM to perform analog MIMO beamforming. These efforts can be broadly categorized into two categories according to the system architecture.
\begin{itemize} 
\item {\emph{Fully-Analog Architecture:}} In \cite{JSAC_2023_An_Stacked}, \emph{An et al.} harnessed a pair of SIM devices to automatically perform MIMO precoding and combining as EM waves propagate through them. The aim is to minimize the error between the actual end-to-end channel matrix and a target diagonal one representing parallel interference-free subchannels. To achieve this, a gradient descent algorithm was developed to determine the phase shifts for all the metasurface layers. Moreover, they explored the integration of SIM into the radome of a BS to enable multiuser downlink communications \cite{ICC_2023_An_Stacked, arXiv_2023_An_Stacked}. As shown in Fig. \ref{fig_5}, thanks to analog beamforming in the wave domain, the signal for each data stream can be directly radiated from the corresponding transmit antenna, allowing the communication system to work with lower-resolution digital-to-analog converters (DACs) and fewer RF chains. Additionally, \emph{Darsena et al.} \cite{arXiv_2024_Darsena_Design} enhanced channel capacity by inserting active layers with amplifier chips to tune both phase and amplitude. The authors of \cite{WCL_2024_Papazafeiropoulos_Achievable} used statistical CSI to optimize the phase shifts of the SIM for maximizing the sum rate in multiuser downlink scenarios. By contrast, \emph{Rezvani et al.} \cite{arXiv_2024_Rezvani_Uplink} investigated a SIM-aided uplink communication scenario, proposing an interior point method to maximize the sum rate while accounting for hardware imperfections. The results showed that SIM can outperform digital phased arrays with the same number of RF chains under realistic 3GPP channels. More recently, \emph{Nerini er al.} \cite{CL_2024_Nerini_Physically} explored the impact of mutual coupling among meta-atoms in SIM, while \cite{arXiv_2024_Papazafeiropoulos_Performance} placed a SIM in the environment to enhance the capability of a single-layer RIS for shaping the wireless channel. Additionally, the fully-analog wideband beamforming design was investigated in \cite{APWCS_2024_Li_Stacked}, where the influence of the number of subcarriers and transmission bandwidth were also characterized.
\item {\emph{Hybrid Architecture:}} While SIM can perform analog signal processing in the wave domain, it is important to note that a larger surface aperture is required for handling more data streams, which may limit its practical scalability. To address this issue, the hybrid architecture that combines SIM with low-complexity digital signal processors has recently gained attention. Specifically, \emph{Papazafeiropoulos et al.} \cite{TWC_2024_Papazafeiropoulos_Achievable} designed a hybrid architecture harnessing two SIMs to enhance transmit precoding and receiver combining in HMIMO communications. The phase shifts of SIMs at the transceivers and the transmit covariance matrix were iteratively optimized using the projected gradient method. Furthermore, \emph{Perovi\'c} \cite{CL_2024_Perovic_Mutual} leveraged the cutoff rate as a proxy for mutual information optimization. They proposed an alternating projected gradient method to maximize the channel cutoff rate of a SIM-based HMIMO system by adjusting the transmit precoding and phase shifts of transmitting and receiving SIMs layer-by-layer. Additionally, they investigated the energy efficiency of SIM-aided MIMO broadcast systems \cite{arXiv_2024_Perovic_Energy}. The uplink-downlink duality was utilized to simplify the energy efficiency maximization problems formulated for dirty paper coding (DPC) and linear precoding, while the transmit covariance matrices were optimized using the successive convex approximation technique. Moreover, they demonstrated that the system energy efficiency can be improved by optimizing the distribution of meta-atoms across layers.
\end{itemize}

\begin{table*}[]
\renewcommand\arraystretch{1.25}
\centering
\caption{Recent contributions of applying SIM in wireless communication systems}
\begin{tabular}{l|l|l|l|l|l|l}
\hline\hline
Ref. & Year & Direction & Task & Objective function & Optimization method & Feature \\ \hline\hline
\cite{JSAC_2023_An_Stacked} & 2023 & \emph{n/a} & MIMO beamforming & Mean square error & Projected gradient descent & \emph{n/a} \\ \hline
\cite{TWC_2024_Papazafeiropoulos_Achievable} & 2024 & \emph{n/a} & MIMO beamforming & Channel capacity & Projected gradient ascent & Hybrid architecture \\ \hline
\cite{CL_2024_Perovic_Mutual} & 2024 & \emph{n/a} & MIMO beamforming & Channel cutoff rate & Projected gradient ascent & Hybrid architecture \\ \hline\hline
\cite{arXiv_2023_An_Stacked} & 2023 & Downlink & Multiuser precoding & Sum rate & Alternating optimization & Discrete phase shifts \\ \hline
\cite{WCL_2024_Papazafeiropoulos_Achievable} & 2024 & Downlink & Multiuser precoding & Sum rate & Alternating optimization & Statistical CSI \\ \hline
\cite{arXiv_2024_Darsena_Design} & 2024 & Downlink & Multiuser precoding & Sum rate & Block coordinate descent & User selection \\ \hline
\cite{WCL_2024_Lin_Stacked} & 2024 & Downlink & Multiuser precoding & Sum rate & Projected gradient ascent & Antenna selection, user grouping \\ \hline
\cite{arXiv_2024_Liu_Multi} & 2024 & Downlink & Multiuser precoding & Sum rate & Deep reinforcement learning & \emph{n/a} \\ \hline
\cite{arXiv_2024_Yang_Joint} & 2024 & Downlink & Multiuser precoding & Sum rate & Twin-delayed DDPG & \emph{n/a} \\ \hline
\cite{arXiv_2024_Amirhosein_Meta} & 2024 & Downlink & Multiuser precoding & Sum rate & Meta-learning & \emph{n/a} \\ \hline
\cite{WCL_2024_Papazafeiropoulos_Near} & 2024 & Downlink & Multiuser precoding & Weighted sum rate & Block coordinate descent & Near-field communications \\ \hline
\cite{arXiv_2024_Perovic_Energy} & 2024 & Downlink & Multiuser precoding & Energy efficiency & Alternating optimization & Dirty paper coding \\ \hline
\cite{VTC_2024_Jia_Stacked} & 2024 & Downlink & Multiuser precoding & Mean square error & Projected gradient descent & EM response modeling \\ \hline
\cite{arXiv_2024_Rezvani_Uplink} & 2024 & Uplink & Multiuser combining & Sum rate & Projected gradient ascent & EM response modeling \\ \hline
\cite{arXiv_2024_Papazafeiropoulos_Performance} & 2024 & Uplink & Multiuser combining & Sum rate & Projected gradient ascent & Environmental SIM \\ \hline\hline
\cite{APWCS_2024_Yao_Sparse} & 2024 & Uplink & Channel estimation & Sparse recovery & Orthogonal matching pursuit & Sparsity-agnostic \\ \hline
\cite{WCL_2024_Yao_Channel} & 2024 & Uplink & Channel estimation & Mean square error & Random codebook & \emph{n/a} \\ \hline
\cite{WCNC_2024_Nadeem_Hybrid} & 2024 & Uplink & Channel estimation & Mean square error & Projected gradient descent & \emph{n/a} \\ \hline\hline
\cite{TCOM_2024_Li_Stacked} & 2024 & Uplink & CF multiuser combining & SINR & Alternating optimization & \emph{n/a} \\ \hline
\cite{arXiv_2024_Shi_Harnessing} & 2024 & Uplink & CF multiuser combining & Sum rate & Alternating optimization & \emph{n/a} \\ \hline
\cite{arXiv_2024_Shi_Joint} & 2024 & Uplink & CF multiuser combining & Sum rate & Alternating optimization & Antenna-user association \\ \hline\hline
\cite{WCL_2024_Niu_Stacked} & 2024 & Downlink & ISAC beamforming & Sum rate & Projected gradient ascent & A single user \\ \hline
\cite{arXiv_2024_Li_Transmit} & 2024 & Downlink & ISAC beamforming & Mean square error & Projected gradient descent & Dual functions \\ \hline
\cite{arXiv_2024_Wang_Multi} & 2024 & Downlink & ISAC beamforming & Cram\'er–Rao bound & Alternating optimization & Environmental SIM \\ \hline\hline
\cite{arXiv_2024_Huang_Stacked} & 2024 & Downlink & Semantic encoding & Cross entropy & Mini-batch gradient descent & \emph{n/a} \\ \hline
\cite{APWCS_2024_Li_Stacked} & 2024 & \emph{n/a} & Wideband beamforming & Mean square error & Projected gradient descent & \emph{n/a} \\ \hline
\cite{arXiv_2023_An_Stacked} & 2024 & \emph{n/a} & DOA estimation & Mean square error & Projected gradient descent & Wave-domain DFT operator \\ \hline
\cite{OJCOMS_2024_Hassan_Efficient} & 2024 & \emph{n/a} & Hologram generation & Mean square error & Successive refinement & Discrete phase shifts \\ \hline
\cite{CL_2024_Nerini_Physically} & 2024 & \emph{n/a} & Channel enhancement & Channel gain & Successive refinement & Mutual coupling among elements \\ \hline
\cite{APWCS_2024_Niu_Enhancing} & 2024 & \emph{n/a} & Waveform design for PLS & Mean square error & Alternating optimization & Closed-form solution \\ \hline\hline
\end{tabular}
\label{tab2}
\end{table*}
\subsubsection{\bf{Channel Estimation}}
CSI acquisition in SIM-aided communication systems presents a major challenge since the BS has fewer RF chains than the number of meta-atoms on each metasurface layer, which determines the channel dimension that needs to be probed. To tackle this challenge, \emph{Yao et al.} \cite{WCL_2024_Yao_Channel} investigated channel estimation for SIM-assisted multi-user HMIMO communication systems. They collected multiple copies of uplink pilot signals that propagate through the SIM and used the array geometry to identify the subspace that covers any spatial correlation matrices. By leveraging partial information about channel statistics, two subspace-based channel estimators were proposed. They also modeled the channel estimation problem in SIM-assisted mmWave communication systems as a sparse recovery problem and applied an improved orthogonal matching pursuit (OMP) algorithm to estimate channel parameters \cite{APWCS_2024_Yao_Sparse}. Additionally, \emph{Nadeem et al.} \cite{WCNC_2024_Nadeem_Hybrid} developed a novel hybrid-domain channel estimator. The phase shifts of the meta-atoms were optimized using a gradient descent algorithm to minimize the mean square error (MSE) of channel estimates. Simulation results demonstrated that a six-layer SIM with four RF chains achieved the same estimation accuracy as a fully digital channel estimator with $64$ RF chains.

\subsubsection{\bf{ML-based Design}}
As previously mentioned, a number of optimization techniques such as convex optimization, gradient descent, and alternating optimization algorithms have been used to fine-tune the transmission coefficients of SIM for achieving wave-domain signal processing. Nevertheless, the wireless communication environment in practical SIM-aided networks is highly dynamic. Additionally, user feedback is often limited and resource-intensive, degrading the performance of conventional optimization approaches. Against this background, ML techniques have gained attention for their powerful learning abilities. For instance, \emph{Liu et al.} \cite{ICC_2024_Liu_DRL, arXiv_2024_Liu_Multi} developed a customized deep reinforcement learning (DRL) approach to jointly optimize the SIM's phase shifts and transmit power allocation for a downlink multiuser system by continuously learning from simulated wireless environments. A whitening process was utilized to enhance the DRL's robustness. Furthermore, \emph{Yang et al.} \cite{arXiv_2024_Yang_Joint} proposed a joint optimization method for SIM phase shift configuration and power allocation based on the twin delayed deep deterministic policy gradient (TD$3$) algorithm, which achieves a higher sum rate compared to the deep deterministic policy gradient (DDPG) algorithm. They also found that increasing the number of metasurface layers yields diminishing returns. Additionally, the authors of \cite{arXiv_2024_Amirhosein_Meta} investigated the interplay between RIS and SIM. They formulated a resource allocation problem aimed at maximizing the achievable rate while ensuring quality-of-service (QoS) requirements for terminals. By modeling the problem as a Markov decision process, a TD$3$ policy gradient agent was utilized to optimize the system's decision variables, with meta-learning employed to adapt to user mobility.

\subsection{Promising Applications}\label{sec3_d}
In this section, we explore the integration of SIM with emerging communication technologies. Table \ref{tab2} summarizes the existing research efforts on applying SIM in wireless communication systems.

\subsubsection{\bf{Cell-Free (CF) Massive MIMO}}
The user-centric CF network is a promising network architecture that effectively eludes inter-cell interference in conventional cellular networks \cite{IoTJ_2024_Xu_Algorithm, Proc_2024_Shi_RIS}. However, as the number of access points (APs) deployed increases, the CF network faces higher hardware costs, energy consumption, and backhaul overhead. To tackle these issues, \emph{Li et al.} \cite{TCOM_2024_Li_Stacked} explored integrating SIMs into APs to improve the spectrum and energy efficiency of the uplink CF systems. The transmission coefficients of the SIM and the local receiver combiner at each AP were optimized based on the local CSI to maximize the composite channel gain. Subsequently, the central processing unit (CPU) combines the local detection results from all APs to recover the user symbol. The weight vector was designed based on the minimum mean square error (MMSE) criterion, taking into account hardware impairments. Additionally, \emph{Shi et al.} \cite{APWCS_2024_Shi_Uplink, arXiv_2024_Shi_Harnessing} derived a closed-form expression for the spectrum efficiency of a SIM-enhanced CF massive MIMO uplink system, based on which a wave-domain beamforming scheme and a max-min power control algorithm were designed based on the statistical CSI. Furthermore, a two-stage protocol was developed to maximize the SIM-aided CF network capacity \cite{arXiv_2024_Shi_Joint}, where a greedy antenna-user association algorithm was also proposed.

\subsubsection{\bf{Near-Field Communications (NFC)}}
A metasurface typically consists of a large number of low-cost meta-atoms, forming an extremely large-scale surface aperture. Additionally, as future wireless communications are likely to operate at higher frequencies, NFC is becoming mainstream. In \cite{WC_2024_An_Near}, \emph{Jia et al.} \cite{VTC_2024_Jia_Stacked} investigated the SIM-based transceiver design for a multiuser MISO NFC system, where a SIM is integrated into the BS to replace the conventional digital baseband architecture. The spherical wavefront characteristics were leveraged to accurately characterize the near-field channel. Numerical results demonstrated that the SIM can effectively mitigate interference between users, while significantly reducing the processing delay. Moreover, the authors of \cite{WCL_2024_Papazafeiropoulos_Near} maximized the weighted sum rate of multiple users in the near field, where they developed a block coordinate descent algorithm to adjust the transmit power and phase shifts of the SIM.

\subsubsection{\bf{Low-Earth Orbit (LEO) Satellite}}
LEO satellites have gained increasing attention as a crucial supplement to terrestrial wireless networks due to their wide coverage area \cite{IoTJ_2023_Xu_OTFS}. Motivated by this, \emph{Lin et al.} \cite{WCL_2024_Lin_Stacked} suggested installing a lightweight and energy-efficient SIM on a LEO satellite to enable analog beamforming towards multiple users. This would significantly reduce the processing delay and computing load of the satellite compared to traditional digital beamforming methods. Additionally, a user grouping method and an antenna selection algorithm were proposed to further enhance the system's performance. In \cite{arXiv_2024_Liu_Stacked}, \emph{Liu et al.} explored the use of an aerial SIM for monitoring natural disasters. The geomorphic image is first encoded into the transmission coefficients of the SIM's input layer, while the remaining layers construct an EMNN to extract information and send the recognition result to a ground receiving station.

\subsubsection{\bf{Physical Layer Security (PLS)}}
The inherent broadcast property of wireless channels poses significant challenges to PLS. Fortunately, SIMs provide an efficient solution to mitigate the risk of information leakage in undesired directions without requiring excessive RF chains. In \cite{APWCS_2024_Niu_Enhancing}, \emph{Niu et al.} integrated a SIM with a single-antenna transmitter to enable joint modulation, beamforming, and artificial noise generation. The ANN-like structure of the SIM allows it to automatically convert the input carrier signal into the desired output signal. Specifically, a fitting problem between the actual and desired output signals is formulated. An alternating optimization algorithm was then developed to iteratively determine each phase shift and transmit power using closed-form expressions derived. Simulation results confirmed the effectiveness of SIM in achieving PLS with low complexity.

\subsubsection{\bf{ISAC}}
ISAC is emerging as a key technology to address the growing issue of spectrum congestion and to meet the increasing demand for ubiquitous sensing and communication \cite{TWC_2023_An_Fundamental}. The SIM technology is expected to produce a dual-function beam pattern that allows for simultaneous communication with multiple users and the detection of radar targets. For instance, \emph{Niu et al.} \cite{WCL_2024_Niu_Stacked} explored the potential of ISAC precoding in the wave domain. They formulated an optimization problem to maximize system spectrum efficiency while ensuring the power constraint in the desired direction. By constructing a penalty term to account for the sensing power constraint, a customized gradient ascent algorithm was developed to design the phase shifts of the SIM and power allocation at the BS. Additionally, \emph{Li et al.} \cite{arXiv_2024_Li_Transmit} proposed a dual-normalized differential gradient descent algorithm to solve the same problem. It was demonstrated that the wave-domain ISAC precoder can automatically generate a desired beam pattern for the sensing task as multiple data streams transmit through the SIM. Furthermore, \emph{Wang et al.} \cite{arXiv_2024_Wang_Multi} fine-tuned both the transmit beamforming at the BS and the transmission coefficient matrix of the SIM to minimize the Cram\'er-Rao bound (CRB) for parameter estimation, while meeting minimum signal-to-interference-plus-noise ratio (SINR) requirements for communication users.

\begin{figure}[!t]
\centerline{\includegraphics[width = 0.9\columnwidth]{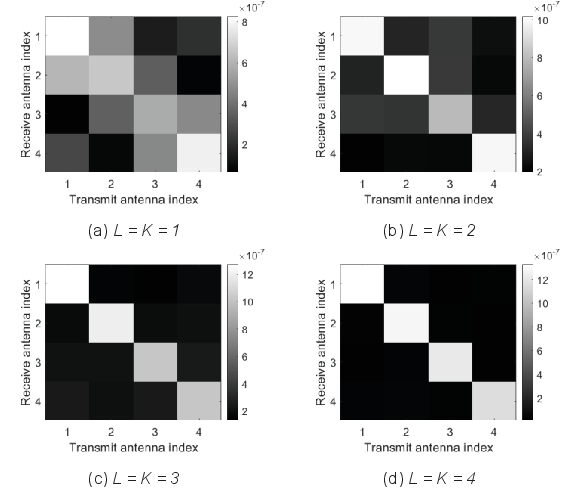}}
\caption{The visualization of the end-to-end MIMO channel, where two SIMs are utilized to perform analog precoding and combining in the wave domain.}
\label{fig_6}
\end{figure}
\begin{figure}[!t]
\centerline{\includegraphics[width = 0.9\columnwidth]{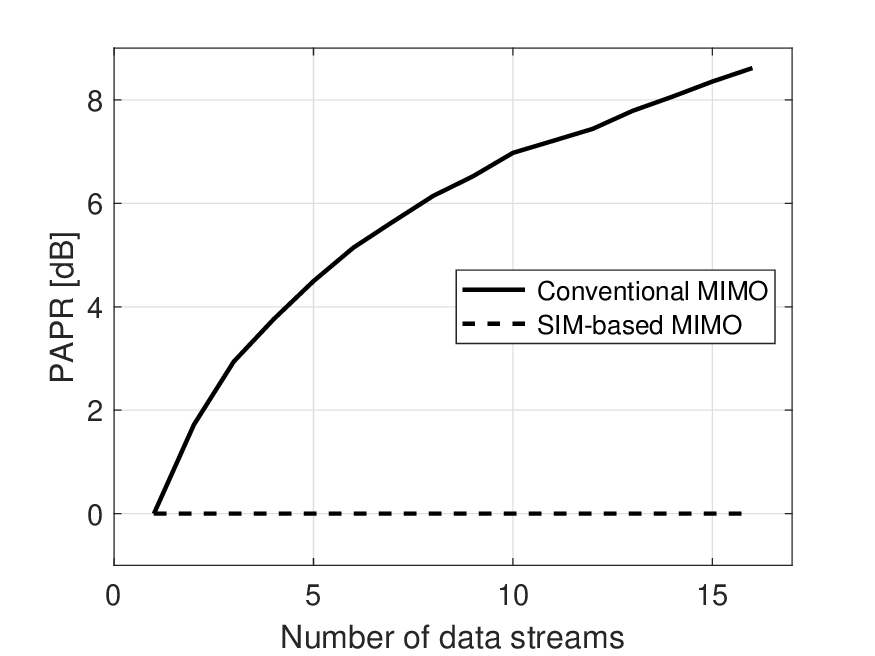}}
\caption{PAPR versus the number of data streams, where binary phase-shift keying modulation is considered.}\label{fig_7}
\end{figure}

\begin{figure*}[!t]
\centerline{\includegraphics[width = 1.9\columnwidth]{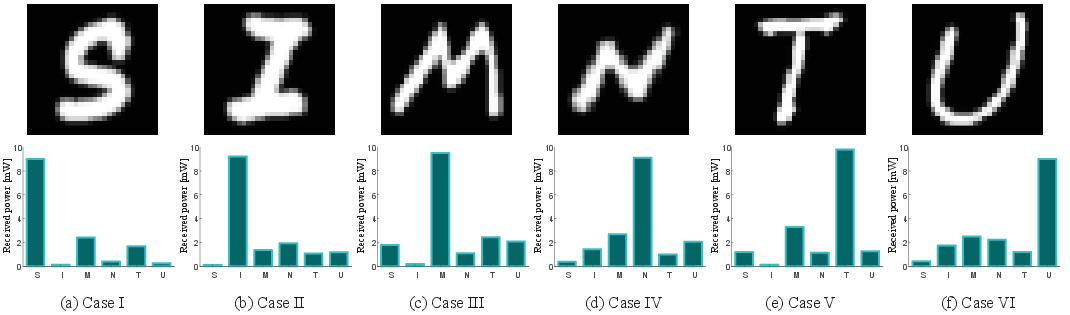}}
\caption{SIM is utilized for semantic encoding in an image classification task-oriented semantic communication system, where the input image and the corresponding energy distribution at the receiving array are shown for six cases.}
\label{fig_8}
\end{figure*}
\subsubsection{\bf{DOA Estimation}}
Accurate DOA information is crucial for developing signal processing algorithms used in wireless communication, radar sensing, navigation, and other applications. However, existing DOA estimation methods typically require RF circuits to sample and demodulate multichannel signals, as well as complex baseband signal processing. This fundamentally limits the processing speed and energy efficiency. In \cite{ICC_2024_An_Stacked, JSAC_2024_An_Two}, a SIM was utilized to estimate the 2D DOA of radio signals. In contrast to conventional designs, a SIM placed in front of a receiver array can be designed to compute the 2D discrete Fourier transform (DFT) of incoming signals. As a result, the receiver array can directly observe the angular spectrum of incoming signals and estimate the DOA by simply probing for the energy peak across the receiver array. This approach eliminates the need for power-inefficient RF chains at each antenna. To achieve this goal, a customized gradient descent algorithm was used to iteratively update the phase shift of each meta-atom in the SIM to minimize the error between the SIM's transfer function and the ideal 2D DFT matrix. Numerical simulations have confirmed that an optimized SIM can perform DOA estimation with remarkable speed and accuracy, achieving an MSE of $-40$ dB. Recently, \emph{Gao et al.} \cite{LSA_2024_Gao_Super} utilized a spatial-temporal multiplexing strategy to achieve high-resolution DOA estimation over a wide field of view. They demonstrated that the angular resolution of the SIM can even exceed the diffraction limit. Additionally, a novel ISAC scenario was demonstrated, where a SIM was utilized to provide DOA estimates for a co-located RIS.

\subsubsection{\bf{Semantic Communications}}
Semantic communications can significantly enhance the efficiency and reliability of information transmission. In \cite{arXiv_2024_Huang_Stacked}, \emph{Huang et al.} demonstrated the potential of employing a SIM to achieve semantic encoding for image recognition tasks. In contrast to traditional communication systems, which transmit modulated signals carrying image data or compressed semantic information, the input layer of the SIM is used for source encoding. The remaining layers form an EMNN, which transforms the signals passing through the input layer into a unique beam towards a receiving antenna corresponding to the image class. Both the source encoding and semantic encoding occur automatically as the carrier EM waves propagate through the SIM. At the receiver, the image is recognized by measuring the signal magnitude across the receiving array. After training the transmission coefficients of meta-atoms, the SIM achieved over $90$\% accuracy in image recognition \cite{arXiv_2024_Huang_Stacked}. Moreover, combining a SIM with conventional amplitude/phase modulation is anticipated to enable multimodal semantic information transmission.

\subsubsection{\bf{Hologram Generation}}
Leveraging SIMs to generate holograms has great potential for various computational imaging applications. To this end, \emph{Hassan et al.} \cite{OJCOMS_2024_Hassan_Efficient} formulated an optimization problem to produce a desired radiation hologram on a 2D plane at a certain distance from the SIM's output layer. They utilized a gradient descent algorithm for SIMs with continuously adjustable phase shifts and an alternating optimization algorithm for SIMs with discrete phase shifts. Simulation results showed that over $90$\% of the radiated power could be concentrated in the desired area using only three metasurface layers with continuous-valued transmission coefficients. However, when using discrete-valued transmission coefficients with two phase shifts, nine metasurface layers are required to achieve the same concentration level. In \cite{TAP_2024_Jia_High}, \emph{Jia et al.} designed a tunable SIM using metasurfaces with a transmission efficiency of $96$\%. By optimizing the binary transmission state of each meta-atom, the SIM can simultaneously achieve multichannel information transmission and holographic image generation.

\begin{figure*}[!t]
\centerline{\includegraphics[width = 1.9\columnwidth]{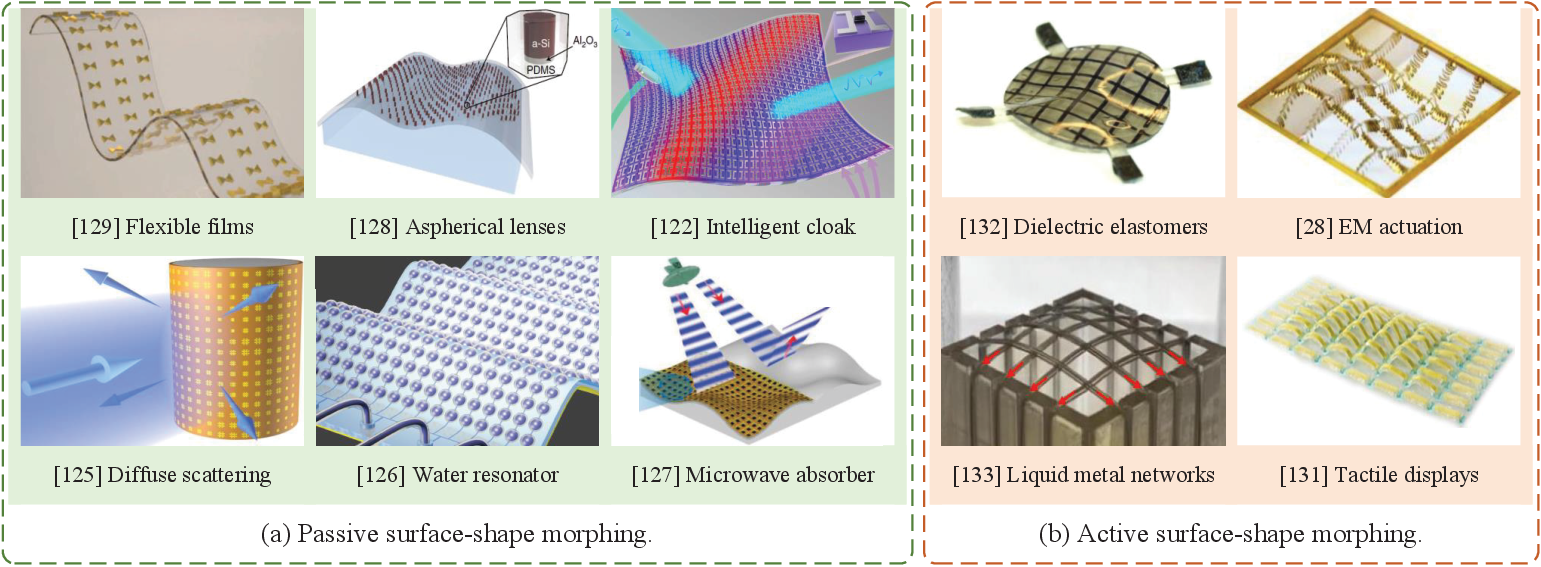}}
\caption{Illustration of several existing FIM prototypes.}
\label{fig_9}
\end{figure*}
\subsection{Case Studies}\label{sec3_e}
In this section, we provide two examples to show the capability of SIM in wireless communications.
\subsubsection{\bf{MIMO Beamforming}}
In Fig. \ref{fig_6}, we evaluate the ability of two SIMs to perform matrix operations for MIMO precoding and combining. Each metasurface layer is comprised of $100$ meta-atoms with a half-wavelength spacing. Moreover, we consider four data streams and use the same simulation parameters as in \cite{JSAC_2023_An_Stacked}. Fig. \ref{fig_6}(a) reveals that when using two SIMs with a single metasurface each, it is hard to form a diagonal channel matrix from the source to the destination. Therefore, each data stream experiences severe interference from other streams. However, as the number of metasurface layers increases, the SIMs gain a stronger inference capability to form multiple parallel subchannels in the physical space. Notably, when each SIM has four metasurface layers, an almost perfectly diagonal channel matrix is formed to enable spatial multiplexing, as seen in Fig. \ref{fig_6}(d).

Moreover, since MIMO precoding and combing are implemented in the wave domain, each data stream can be directly transmitted from the corresponding antenna. This significantly reduces the peak-to-average power ratio (PAPR) of the waveforms, enhancing power amplifier efficiency. In Fig. \ref{fig_7}, we compare the PAPR of transmit waveforms in SIM-based and conventional MIMO systems, considering binary phase shift keying signals and Rayleigh fading. The results indicate that the PAPR of the waveforms in fully digital MIMO systems increases with the number of data streams, as the superimposed signal is transmitted. For example, in conventional MIMO systems, the PAPR exceeds $8$ dB for transmitting $16$ data streams. By contrast, the PAPR of signals in SIM-based MIMO systems remains at $0$ dB due to its favorable constant-envelop property.

\subsubsection{\bf{Semantic Encoding}}
Next, we use the EMNIST dataset to evaluate the performance of the SIM in enabling semantic communications \cite{IJCNN_2017_Cohen_EMNIST}. In contrast to conventional counterparts, the SIM is deployed at the transmitter as a semantic encoder that transforms image information into a unique beam. The system operates at $28$ GHz and features a four-layer SIM, with each layer containing $21 \times 21 = 441$ meta-atoms. The thickness of the SIM is set to $10$ wavelengths. Additionally, the transmit power and the noise power are set to $40$ dBm and $-104$ dBm, respectively. Other simulation setups, such as wireless channels and array patterns, are detailed in \cite{arXiv_2024_Huang_Stacked}, and cross-entropy is utilized as a metric to train the SIM. Fig. \ref{fig_8} illustrates the energy distribution at the receiving array for each input image. For simplicity, we focus on six letters: `S', `I', `M', `N', `T', and `U'. The results demonstrate that the SIM is capable of directing the information-bearing signals to the corresponding antenna. As a result, the image class can be readily recognized by detecting the energy peak, which significantly reduces the hardware cost while improving the processing speed.

\subsection{Open Challenges}\label{sec3_f}
\subsubsection{\bf{Nonlinear Activation Function}}
Although SIM offers advantages in terms of power efficiency and computation speed, its inference capability is limited by the inherent linear transfer function. To overcome this limitation, an effective solution is to cascade a SIM with an electronic neural network for creating a HOENN \cite{JSTQE_2020_Mengu_Analysis}. In HOENN, the power-efficient SIM can preprocess spatial EM waves, reducing the scale requirements and the computation load for the electronic neuron network \cite{arXiv_2024_Liu_Stacked}. Moreover, by extracting the key information from the phase characteristics of the incident EM waves and transforming them into amplitude characteristics, the system can use power-efficient envelope detectors at the receiver to enable the electronic neuron network. As a result, HOENN significantly decreases power consumption and hardware complexity while improving the inference capability of the SIM. In addition, it is possible to integrate nonlinear components, e.g., adjustable varactors, into the meta-atom to replicate the nonlinear activation functions in traditional neural networks, thus implementing various mathematical functions and machine learning tasks \cite{NE_2023_Gao_Programmable}. However, the precise modeling and device design with benign nonlinear response need to be further investigated \cite{NP_2024_Yildirim_Nonlinear}.
\subsubsection{\bf{Performance Evaluation}}
The advent of SIM technology has opened up new possibilities to utilize EM waves in data processing. In essence, SIM enables a more energy-efficient integration of wireless communication and computing by shifting signal processing from the baseband domain to the EM domain, paving the way for future innovations. However, the effective theoretical framework is crucial for characterizing the theoretical limit of SIM. Several key methods for characterizing the relationship between structure and function in EM systems include semi-classical analytical models, heuristic optimization techniques, and topology optimization approaches \cite{APS_2021_Luo_Electromagnetic}. In addition, it requires further investigation to characterize the fundamental tradeoff between the power loss of EM signals as they pass through the multiple metasurface layers and the benefit of increased depth.

\section{Flexible Intelligent Metasurface (FIM)}
Conventional metasurfaces are typically flat and rigid, which restricts their practical applications. To overcome this limitation, researchers are exploring novel methods for designing and constructing FIMs that can morph into various shapes \cite{JO_2020_Zang_Reconfigurable, AN_2024_Zhou_Flexible}. In this section, we delve into the emerging FIM technology as well as its promising applications in wireless communications. Specifically, we first review the existing hardware prototypes of FIM in Section \ref{sec4_a}. Then, the functionalities of FIM for wireless communications are discussed in Section \ref{sec4_b} and some numerical results are provided in Section \ref{sec4_c}. Finally, Section \ref{sec4_d} presents the practical challenges when deploying FIM in wireless networks to motivate future research.

\begin{figure}[!t]
\centerline{\includegraphics[width = 0.9\columnwidth]{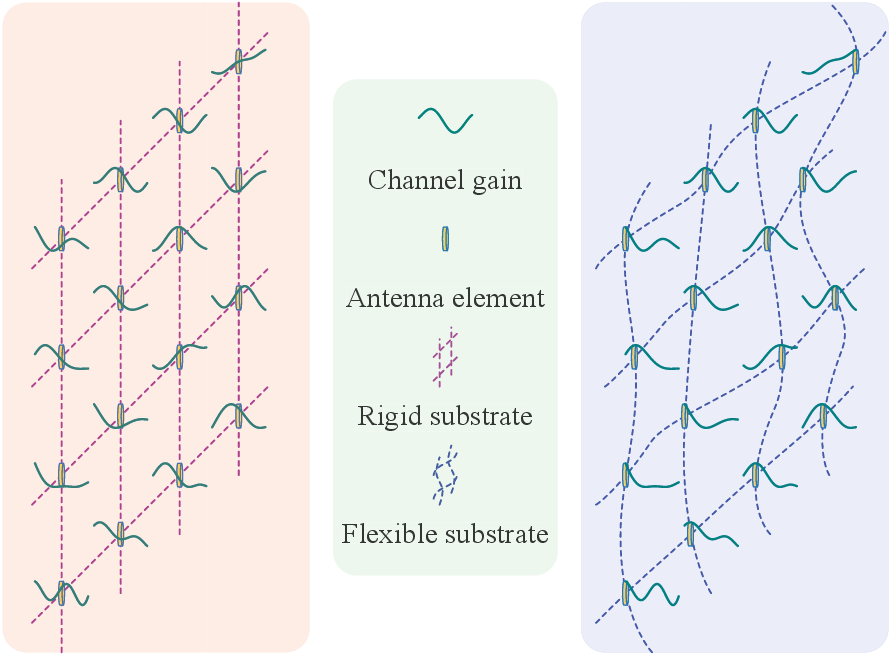}}
\caption{Comparison between FIM and conventional rigid array. Note that by beneficially morphing its surface shape, each radiating element on FIM can attain the highest channel gain.}
\label{fig_10}
\end{figure}
\subsection{Hardware Prototypes}\label{sec4_a}
As illustrated in Fig. \ref{fig_9}, FIMs can be divided into two categories based on their ability to morph their surface shape actively:
\subsubsection{\bf{Passive Morphing}}
FIMs can be conformally deployed on curved surfaces to alter their EM properties, which show great potential in EM camouflage and stealth applications \cite{TAP_2022_Benoni_Planning, ACS_2017_Cong_Perovskite, NP_2020_Qian_Deep, Small_2023_Nam_Flexible}. For example, by randomly arranging metallic patches, the authors of \cite{SR_2016_Zhang_Broadband, OE_2016_Zhao_Achieving} showed that FIMs can significantly reduce specular reflection at various angles of incidence. This diffuse scattering property remains effective, even when the FIM is wrapped around a curved surface. Measurements indicated that coating a cylindrical metal surface with an FIM can reduce the radar cross-section by $10$ dB across the \emph{X} band \cite{OE_2016_Zhao_Achieving}. Furthermore, \emph{Song et al.} \cite{AOM_2017_Song_Water} developed a soft, water-resonator-based FIM that functions as an active absorber across \emph{Ku}, \emph{K}, and \emph{Ka} bands, achieving near-perfect absorption of $99$\% even when the FIM is bent into different curvatures. More recently, \emph{Wang et al.} \cite{AM_2021_Wang_Pangolin} designed a stretchable FIM that exhibits robust microwave‐absorbing capacity under stretching. Additionally, \cite{NC_2016_Kamali_Decoupling} demonstrated an aspherical lens by covering a cylindrical lens with an FIM, and \cite{AM_2011_Aksu_Flexible} showed that the EM responses of FIMs can be actively tuned by mechanically stretching the flexible substrate. Motivated by catenary optics, \emph{Huang et al.} \cite{AS_2019_Huang_Catenary} developed a flexible substrate-based microfabrication process capable of achieving a minimum feature size of the sub-micrometer scale.
\subsubsection{\bf{Active Morphing}}
Surface shape-morphable FIMs have significant potential for applications such as flexible displays, surgical instruments, and soft robots. However, creating FIMs that can alter their shapes in response to external stimuli after fabrication poses a substantial challenge. In past years, considerable progress has been made to tackle this issue \cite{NC_2024_An_Energy}. For instance, \emph{Hajiesmaili et al.} \cite{NC_2019_Hajiesmaili_Reconfigurable} developed an FIM using specially designed electrodes made of carbon nanotubes sandwiched between thin elastomer sheets. By applying different voltages to specific internal electrodes, the surface shape can be reconfigured through reversible deformation. Moreover, \emph{Ni et al.} \cite{NC_2022_Ni_Soft} presented an FIM with shape-morphing capabilities using liquid metal networks embedded in a flexible material. This design takes advantage of the liquid-solid phase transition of the liquid metal, allowing the surface to rapidly and continuously transform into complex 3D shapes as needed. Similarly, \emph{Bai et al.} \cite{Nature_2022_Bai_A} created an FIM made of a mesh of filamentary metal traces. By programming distributed Lorentz forces generated by passing electrical currents in the presence of a static magnetic field, local deformations can be accurately controlled with high fidelity. Through an integrated digital control scheme and in-situ stereo imaging feedback, the FIM is capable of morphing into a variety of target shapes with response time in milliseconds. Recently, \emph{Wang et al.} \cite{SA_2023_Wang_Passively} proposed an FIM that can change its surface shape on demand using a matrix of ionic actuators.

\begin{figure}[!t]
\centerline{\includegraphics[width = 0.9\columnwidth]{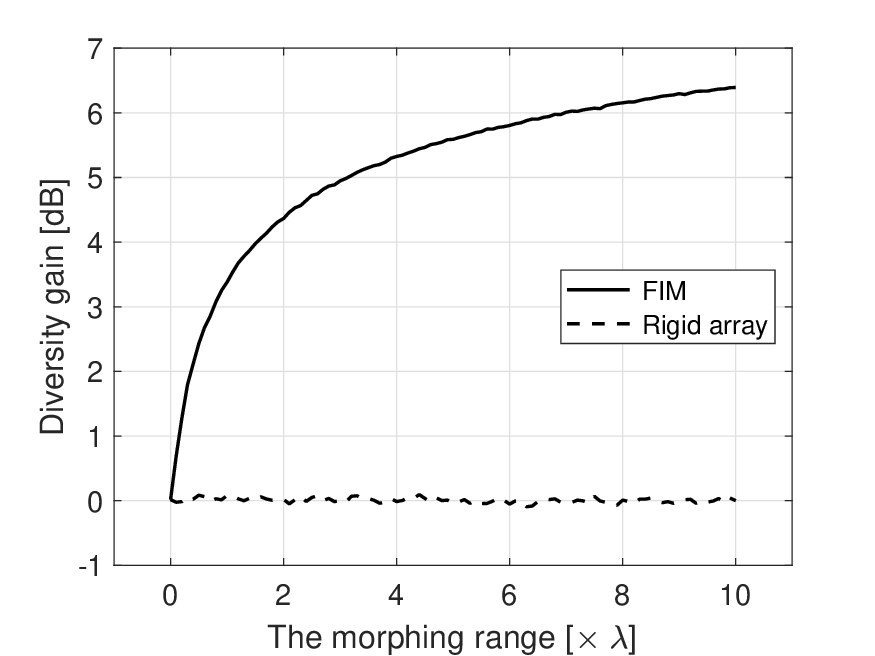}}
\caption{Diversity gain versus the morphing range of the FIM.}
\label{fig_11}
\end{figure}
\begin{figure}[!t]
\centerline{\includegraphics[width = 0.9\columnwidth]{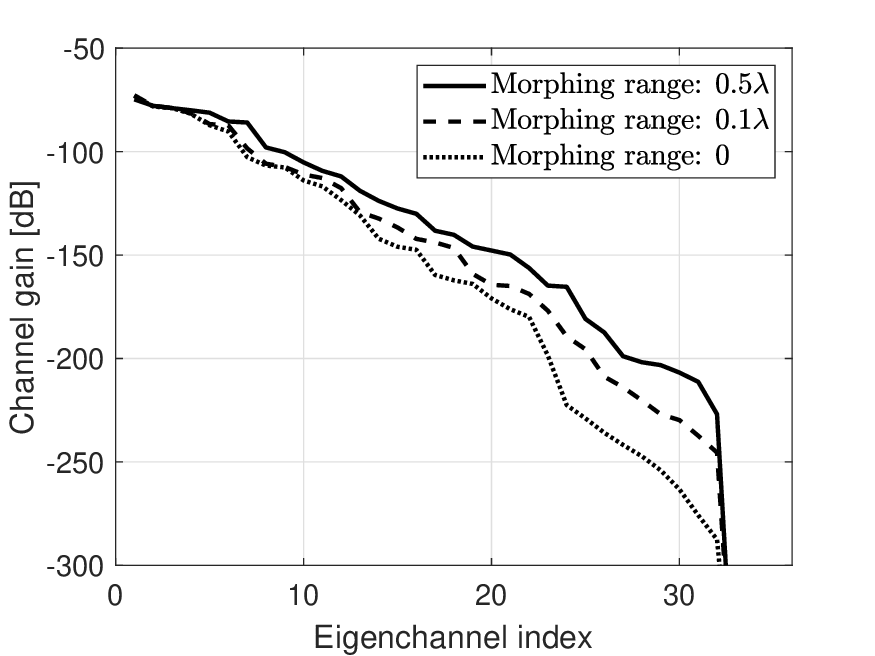}}
\caption{Channel gain versus the eigenchannel index.}
\label{fig_12}
\end{figure}
\subsection{Functionalities of FIM in Wireless Communications}\label{sec4_b}
The unique ability of FIMs to morph their surface shape enables them to potentially enhance signal quality for information transmission. This is particularly beneficial for telecommunications operating at high frequencies, where the channel coherence distance is relatively small. In this subsection, we will discuss recent progress in applying FIM in wireless communications.
\subsubsection{\bf{MIMO Capacity Enhancement}}
In \cite{TCOM_2025_An_Flexible}, \emph{An et al.} explored the use of FIMs as transceiver arrays in point-to-point MIMO communications. They characterized the capacity limits of FIM-aided MIMO transmissions over frequency-flat fading channels, by simultaneously optimizing the 3D surface shapes of the transmitting and receiving FIMs as well as the transmit covariance matrix. The optimization took into account constraints including the transmit power budget and the maximum morphing range of the FIMs. They developed an efficient block coordinate descent algorithm that alternately updates the FIM surface shapes and the transmit covariance matrix, until reaching a local optimum. Numerical results verified that FIMs can achieve higher MIMO capacity than conventional rigid arrays. Specifically, the MIMO channel capacity can even be doubled under certain system configurations.

\subsubsection{\bf{Multiuser Interference Mitigation}}
In practice, an FIM can be integrated with a BS to enhance its ability to mitigate inter-user interference. Specifically, by beneficially morphing the surface shape of the FIM, the effective channels between different users can be designed to be nearly orthogonal to each other. Motivated by this, \emph{An et al.} \cite{GLOBECOM_2025_An_Downlink, TWC_2024_An_Flexible} investigated multiuser downlink communications, aiming to minimize the transmit power at the BS by jointly optimizing transmit beamforming and the FIM surface shape, subject to the QoS requirements for all users and the maximum morphing range of the FIM. When considering a single user, it was found that the optimal 3D surface shape of the FIM can be achieved by independently adjusting each element to the position with the strongest channel gain, as shown in Fig. \ref{fig_10}. For multiuser communication scenarios, the FIM surface shape and the transmit beamformer were alternately optimized. Simulation results showed that the FIM can reduce the required transmit power by about $3$ dB compared to conventional rigid 2D arrays while maintaining the same data rate. Moreover, \emph{Yang et al.} \cite{arXiv_2024_Yang_Flexible} explored the potential of rotatable, bendable, and foldable FIMs for creating customized radiation patterns for serving multiple users.

\subsection{Case Studies}\label{sec4_c}
In this subsection, we provide two case studies to demonstrate the benefits of deploying FIMs in communication systems for improving diversity and multiplexing gains.
\subsubsection{\bf{Diversity Gain}}
We first examine a single-user scenario to demonstrate the diversity gain provided by FIM. For simplicity, we assume correlated Rayleigh fading, where the spatial correlation between different positions of each element is characterized by \emph{Lemma 2} in \cite{JSAC_2020_Pizzo_Spatially}. Fig. \ref{fig_11} shows the diversity gain versus the morphing range. Compared to a conventional rigid array, FIM significantly improves the diversity gain, which increases logarithmically with the morphing range. Specifically, an FIM with a morphing range of $0.8\lambda$ can boost the diversity gain by $3$ dB. However, achieving a $6$dB diversity gain may require a morphing range of $7.2\lambda$, which presents a considerable engineering challenge for the elasticity of the substrate material.
\subsubsection{\bf{Multiplexing Gain}}
In Fig. \ref{fig_12}, we consider a MIMO scenario to evaluate the multiplexing gain. Specifically, a pair of square FIMs, each consisting of $7 \times 7 = 49$ antenna elements, are deployed at the source and destination. Eight scatterers are randomly positioned in the environment, and the path loss is determined using the method described in \cite{TCOM_2025_An_Flexible}. It is observed from Fig. \ref{fig_12} that the FIMs can morph their surface shapes to improve the channel gain of all eigenchannels, particularly those with poor channel quality. With a morphing range of $0.1 \lambda$, the weak eigenchannels are enhanced by over $40$ dB compared to conventional rigid antenna arrays. Increasing the morphing range of FIMs to $0.5 \lambda$ provides a larger space to morph their surface shapes. Consequently, the eigenchannel gain is further increased by about $20$ dB, which means that more eigenchannels can be utilized for data transmission.

\subsection{Open Challenges}\label{sec4_d}
Next, we discuss three major challenges arising from deploying FIMs into wireless networks.
\subsubsection{\bf{Channel Estimation}}
The performance gain of FIM depends heavily on having perfect knowledge of CSI. However, channel estimation in FIM-aided systems is more challenging since it involves estimating channels across a continuous space. Advanced techniques such as compressed sensing and the space-alternating generalized expectation maximization algorithm can be utilized to estimate channels with moderate overhead. Moreover, the surface shapes of FIMs can potentially be optimized to attain even higher channel estimation accuracy, while further investigation is needed to develop detailed protocols.
\subsubsection{\bf{Surface-Shape Morphing}}
Although advanced EM actuation technology provides dynamic flexibility, the surface shape morphing ability of a practical FIM may be hindered by hardware limitations and may result in additional energy consumption. This motivates further research to develop efficient solutions. Moreover, for rapidly time-varying channels with shorter coherence time, it may be beneficial to optimize the ergodic performance over multiple coherence blocks. Designing robust surface shape morphing methods to strike tradeoffs between achievable performance and implementation complexity is also an important research topic.
\subsubsection{\bf{Practical Deployment}}
Integrating FIMs into existing network infrastructures presents considerable challenges. For outdoor applications, FIMs can be incorporated into clothing, parachutes, or hot-air balloons. In indoor environments, FIMs can be seamlessly integrated into curtains, functioning as signal reflectors or APs. However, it is crucial to evaluate the resilience of these conformal materials. By skillfully combining passive and active surface-shape morphing strategies, the durability of FIMs in real-world applications can be further enhanced, while further research is needed to assess their performance and develop effective protocols.

\section{Conclusion}
In this paper, we have provided an overview of emerging intelligent metasurface technologies and their applications in wireless communications. We first examined recent RIS experiments and explored its applications from four perspectives. We also explained the underlying principles of 3D SIM and discussed relevant prototypes. Numerical results were provided to illustrate SIM's potential for analog signal processing. Finally, we reviewed the state-of-the-art of FIM technology, discussing its impact on wireless communication and identifying key challenges of integrating FIMs into wireless networks. We hope that this comprehensive perspective will inspire future research efforts to achieve the grand vision of efficient communication and computing.

\bibliography{ref}

\begin{thebibliography}{100}
\providecommand{\url}[1]{#1}
\csname url@samestyle\endcsname
\providecommand{\newblock}{\relax}
\providecommand{\bibinfo}[2]{#2}
\providecommand{\BIBentrySTDinterwordspacing}{\spaceskip=0pt\relax}
\providecommand{\BIBentryALTinterwordstretchfactor}{4}
\providecommand{\BIBentryALTinterwordspacing}{\spaceskip=\fontdimen2\font plus
\BIBentryALTinterwordstretchfactor\fontdimen3\font minus \fontdimen4\font\relax}
\providecommand{\BIBforeignlanguage}[2]{{%
\expandafter\ifx\csname l@#1\endcsname\relax
\typeout{** WARNING: IEEEtran.bst: No hyphenation pattern has been}%
\typeout{** loaded for the language `#1'. Using the pattern for}%
\typeout{** the default language instead.}%
\else
\language=\csname l@#1\endcsname
\fi
#2}}
\providecommand{\BIBdecl}{\relax}
\BIBdecl

\bibitem{LSA_2014_Cui_Coding}
T.~J. Cui, M.~Q. Qi, X.~Wan, J.~Zhao, and Q.~Cheng, ``Coding metamaterials, digital metamaterials and programmable metamaterials,'' \emph{Light: Science \& Applications}, vol.~3, no.~10, pp. e218--e218, Oct. 2014.

\bibitem{TAP_2022_Barbuto_Metasurfaces}
M.~Barbuto, Z.~Hamzavi-Zarghani, M.~Longhi, A.~Monti, D.~Ramaccia, S.~Vellucci, A.~Toscano, and F.~Bilotti, ``Metasurfaces 3.0: A new paradigm for enabling smart electromagnetic environments,'' \emph{IEEE Trans. Antennas Propag.}, vol.~70, no.~10, pp. 8883--8897, Oct. 2022.

\bibitem{NC_2018_Zhang_Space}
L.~Zhang, X.~Q. Chen, S.~Liu, Q.~Zhang, J.~Zhao, J.~Y. Dai, G.~D. Bai, X.~Wan, Q.~Cheng, G.~Castaldi \emph{et~al.}, ``Space-time-coding digital metasurfaces,'' \emph{Nature Communications}, vol.~9, no.~1, p. 4334, Oct. 2018.

\bibitem{LSA_2019_Ma_Smart}
Q.~Ma, G.~D. Bai, H.~B. Jing, C.~Yang, L.~Li, and T.~J. Cui, ``Smart metasurface with self-adaptively reprogrammable functions,'' \emph{Light: Science \& Applications}, vol.~8, no.~1, p.~98, Oct. 2019.

\bibitem{Science_2011_Yu_Light}
N.~Yu, P.~Genevet, M.~A. Kats, F.~Aieta, J.-P. Tetienne, F.~Capasso, and Z.~Gaburro, ``Light propagation with phase discontinuities: Generalized laws of reflection and refraction,'' \emph{Science}, vol. 334, no. 6054, pp. 333--337, Sep. 2011.

\bibitem{Science_2014_Silva_Performing}
A.~Silva, F.~Monticone, G.~Castaldi, V.~Galdi, A.~Al{\`u}, and N.~Engheta, ``Performing mathematical operations with metamaterials,'' \emph{Science}, vol. 343, no. 6167, pp. 160--163, Jan. 2014.

\bibitem{APM_2012_Holloway_An}
C.~L. Holloway, E.~F. Kuester, J.~A. Gordon, J.~O'Hara, J.~Booth, and D.~R. Smith, ``An overview of the theory and applications of metasurfaces: The two-dimensional equivalents of metamaterials,'' \emph{IEEE Antennas Propag. Mag.}, vol.~54, no.~2, pp. 10--35, Feb. 2012.

\bibitem{JSAC_2020_Renzo_Smart}
M.~Di~Renzo, A.~Zappone, M.~Debbah, M.-S. Alouini, C.~Yuen, J.~de~Rosny, and S.~Tretyakov, ``Smart radio environments empowered by reconfigurable intelligent surfaces: How it works, state of research, and the road ahead,'' \emph{IEEE J. Sel. Areas Commun.}, vol.~38, no.~11, pp. 2450--2525, Nov. 2020.

\bibitem{CST_2021_Liu_Reconfigurable}
Y.~Liu, X.~Liu, X.~Mu, T.~Hou, J.~Xu, M.~Di~Renzo, and N.~Al-Dhahir, ``Reconfigurable intelligent surfaces: Principles and opportunities,'' \emph{IEEE Commun. Surveys Tuts.}, vol.~23, no.~3, pp. 1546--1577, 3rd Quarter 2021.

\bibitem{JSTSP_2022_Pan_An}
C.~Pan, G.~Zhou, K.~Zhi, S.~Hong, T.~Wu, Y.~Pan, H.~Ren, M.~D. Renzo, A.~Lee~Swindlehurst, R.~Zhang, and A.~Y. Zhang, ``An overview of signal processing techniques for {RIS/IRS}-aided wireless systems,'' \emph{IEEE J. Sel. Topics Signal Process.}, vol.~16, no.~5, pp. 883--917, May 2022.

\bibitem{SPM_2022_Bjornson_Reconfigurable}
E.~Björnson, H.~Wymeersch, B.~Matthiesen, P.~Popovski, L.~Sanguinetti, and E.~de~Carvalho, ``Reconfigurable intelligent surfaces: A signal processing perspective with wireless applications,'' \emph{IEEE Signal Process. Mag.}, vol.~39, no.~2, pp. 135--158, Feb. 2022.

\bibitem{CST_2020_Gong_Toward}
S.~Gong, X.~Lu, D.~T. Hoang, D.~Niyato, L.~Shu, D.~I. Kim, and Y.-C. Liang, ``Toward smart wireless communications via intelligent reflecting surfaces: A contemporary survey,'' \emph{IEEE Commun. Surveys Tuts.}, vol.~22, no.~4, pp. 2283--2314, 4th Quarter 2020.

\bibitem{CST_2022_Zheng_A}
B.~Zheng, C.~You, W.~Mei, and R.~Zhang, ``A survey on channel estimation and practical passive beamforming design for intelligent reflecting surface aided wireless communications,'' \emph{IEEE Commun. Surveys Tuts.}, vol.~24, no.~2, pp. 1035--1071, 2nd Quarter 2022.

\bibitem{VTM_2024_Basar_Reconfigurable}
E.~Basar, G.~C. Alexandropoulos, Y.~Liu, Q.~Wu, S.~Jin, C.~Yuen, O.~A. Dobre, and R.~Schober, ``Reconfigurable intelligent surfaces for {6G}: Emerging hardware architectures, applications, and open challenges,'' \emph{IEEE Veh. Technol. Mag.}, vol.~19, no.~3, pp. 27--47, Sep. 2024.

\bibitem{WC_2024_An_Stacked}
J.~An, C.~Yuen, C.~Xu, H.~Li, D.~W.~K. Ng, M.~Di~Renzo, M.~Debbah, and L.~Hanzo, ``Stacked intelligent metasurface-aided {MIMO} transceiver design,'' \emph{IEEE Wireless Commun.}, vol.~31, no.~4, pp. 123--131, Apr. 2024.

\bibitem{arXiv_2024_Liu_Stacked}
H.~Liu, J.~An, X.~Jia, S.~Lin, X.~Yao, L.~Gan, B.~Clerckx, C.~Yuen, M.~Bennis, and M.~Debbah, ``Stacked intelligent metasurfaces for wireless sensing and communication: Applications and challenges,'' \emph{arXiv preprint arXiv:2407.03566}, 2024.

\bibitem{JO_2020_Zang_Reconfigurable}
G.~Zang, Z.~Liu, W.~Deng, and W.~Zhu, ``Reconfigurable metasurfaces with mechanical actuations: towards flexible and tunable photonic devices,'' \emph{J. Optics}, vol.~23, no.~1, p. 013001, Dec. 2020.

\bibitem{AN_2024_Zhou_Flexible}
Y.~Zhou, S.~Wang, J.~Yin, J.~Wang, F.~Manshaii, X.~Xiao, T.~Zhang, H.~Bao, S.~Jiang, and J.~Chen, ``Flexible metasurfaces for multifunctional interfaces,'' \emph{ACS nano}, vol.~18, no.~4, pp. 2685--2707, Jan. 2024.

\bibitem{WC_2024_An_Codebook}
J.~An, C.~Xu, Q.~Wu, D.~W.~K. Ng, M.~Di~Renzo, C.~Yuen, and L.~Hanzo, ``Codebook-based solutions for reconfigurable intelligent surfaces and their open challenges,'' \emph{IEEE Wireless Commun.}, vol.~31, no.~2, pp. 134--141, Apr. 2024.

\bibitem{TWC_2019_Huang_Reconfigurable}
C.~Huang, A.~Zappone, G.~C. Alexandropoulos, M.~Debbah, and C.~Yuen, ``Reconfigurable intelligent surfaces for energy efficiency in wireless communication,'' \emph{IEEE Trans. Wireless Commun.}, vol.~18, no.~8, pp. 4157--4170, Aug. 2019.

\bibitem{TCOM_2021_Wu_Intelligent}
Q.~Wu, S.~Zhang, B.~Zheng, C.~You, and R.~Zhang, ``Intelligent reflecting surface-aided wireless communications: A tutorial,'' \emph{IEEE Trans. Commun.}, vol.~69, no.~5, pp. 3313--3351, May 2021.

\bibitem{TCOM_2022_An_Low}
J.~An, C.~Xu, L.~Gan, and L.~Hanzo, ``Low-complexity channel estimation and passive beamforming for {RIS}-assisted {MIMO} systems relying on discrete phase shifts,'' \emph{IEEE Trans. Commun.}, vol.~70, no.~2, pp. 1245--1260, Feb. 2022.

\bibitem{Science_2018_Lin_All}
X.~Lin, Y.~Rivenson, N.~T. Yardimci, M.~Veli, Y.~Luo, M.~Jarrahi, and A.~Ozcan, ``All-optical machine learning using diffractive deep neural networks,'' \emph{Science}, vol. 361, no. 6406, pp. 1004--1008, Jul. 2018.

\bibitem{NE_2022_Liu_A}
C.~Liu, Q.~Ma, Z.~J. Luo, Q.~R. Hong, Q.~Xiao, H.~C. Zhang, L.~Miao, W.~M. Yu, Q.~Cheng, L.~Li, and T.~J. Cui, ``A programmable diffractive deep neural network based on a digital-coding metasurface array,'' \emph{Nature Electronics}, vol.~5, no.~2, pp. 113--122, Feb. 2022.

\bibitem{JSAC_2023_An_Stacked}
J.~An, C.~Xu, D.~W.~K. Ng, G.~C. Alexandropoulos, C.~Huang, C.~Yuen, and L.~Hanzo, ``Stacked intelligent metasurfaces for efficient holographic {MIMO} communications in {6G},'' \emph{IEEE J. Sel. Areas Commun.}, vol.~41, no.~8, pp. 2380--2396, Aug. 2023.

\bibitem{JSAC_2024_An_Two}
J.~An, C.~Yuen, Y.~L. Guan, M.~Di~Renzo, M.~Debbah, H.~Vincent~Poor, and L.~Hanzo, ``Two-dimensional direction-of-arrival estimation using stacked intelligent metasurfaces,'' \emph{IEEE J. Sel. Areas Commun.}, vol.~42, no.~10, pp. 2786--2802, Oct. 2024.

\bibitem{LSA_2022_Luo_Metasurface}
X.~Luo, Y.~Hu, X.~Ou, X.~Li, J.~Lai, N.~Liu, X.~Cheng, A.~Pan, and H.~Duan, ``Metasurface-enabled on-chip multiplexed diffractive neural networks in the visible,'' \emph{Light: Science \& Applications}, vol.~11, no.~1, p. 158, May 2022.

\bibitem{Nature_2022_Bai_A}
Y.~Bai, H.~Wang, Y.~Xue, Y.~Pan, J.-T. Kim, X.~Ni, T.-L. Liu, Y.~Yang, M.~Han, Y.~Huang \emph{et~al.}, ``A dynamically reprogrammable surface with self-evolving shape morphing,'' \emph{Nature}, vol. 609, no. 7928, pp. 701--708, Sep. 2022.

\bibitem{TWC_2024_An_Flexible}
J.~An, C.~Yuen, M.~D. Renzo, M.~Debbah, H.~V. Poor, and L.~Hanzo, ``Flexible intelligent metasurfaces for downlink multiuser {MISO} communications,'' \emph{IEEE Trans. Wireless Commun.}, vol.~24, no.~4, pp. 2940--2955, 2025.

\bibitem{TCOM_2025_An_Flexible}
J.~An, Z.~Han, D.~Niyato, M.~Debbah, C.~Yuen, and L.~Hanzo, ``Flexible intelligent metasurfaces for enhancing {MIMO} communications,'' \emph{IEEE Trans. Commun.}, pp. 1--15, 2025, Early Access.

\bibitem{Book_2005_Tse_Fundamentals}
D.~Tse and P.~Viswanath, \emph{Fundamentals of Wireless Communication}.\hskip 1em plus 0.5em minus 0.4em\relax Cambridge university press, 2005.

\bibitem{AP_2024_Bilotti_Reconfigurable}
F.~Bilotti, M.~Barbuto, Z.~Hamzavi-Zarghani, M.~Karamirad, M.~Longhi, A.~Monti, D.~Ramaccia, L.~Stefanini, A.~Toscano, and S.~Vellucci, ``Reconfigurable intelligent surfaces as the key-enabling technology for smart electromagnetic environments,'' \emph{Advances in Physics: X}, vol.~9, no.~1, p. 2299543, Jan. 2024.

\bibitem{PRA_2019_Liu_Intelligent}
F.~Liu, O.~Tsilipakos, A.~Pitilakis, A.~C. Tasolamprou, M.~S. Mirmoosa, N.~V. Kantartzis, D.-H. Kwon, J.~Georgiou, K.~Kossifos, M.~A. Antoniades \emph{et~al.}, ``Intelligent metasurfaces with continuously tunable local surface impedance for multiple reconfigurable functions,'' \emph{Physical Review Applied}, vol.~11, no.~4, p. 044024, Apr. 2019.

\bibitem{Access_2020_Dai_Reconfigurable}
L.~Dai, B.~Wang, M.~Wang, X.~Yang, J.~Tan, S.~Bi, S.~Xu, F.~Yang, Z.~Chen, M.~D. Renzo, C.-B. Chae, and L.~Hanzo, ``Reconfigurable intelligent surface-based wireless communications: Antenna design, prototyping, and experimental results,'' \emph{IEEE Access}, vol.~8, pp. 45\,913--45\,923, Aug. 2020.

\bibitem{TCOM_2021_Pei_RIS}
X.~Pei, H.~Yin, L.~Tan, L.~Cao, Z.~Li, K.~Wang, K.~Zhang, and E.~Björnson, ``{RIS}-aided wireless communications: Prototyping, adaptive beamforming, and indoor/outdoor field trials,'' \emph{IEEE Trans. Commun.}, vol.~69, no.~12, pp. 8627--8640, Dec. 2021.

\bibitem{TAP_2021_Huang_Active}
C.~Huang, B.~Zhao, J.~Song, C.~Guan, and X.~Luo, ``Active transmission/absorption frequency selective surface with dynamical modulation of amplitude,'' \emph{IEEE Trans. Antennas Propag.}, vol.~69, no.~6, pp. 3593--3598, Jun. 2021.

\bibitem{OJCOMS_2022_Trichopoulos_Design}
G.~C. Trichopoulos, P.~Theofanopoulos, B.~Kashyap, A.~Shekhawat, A.~Modi, T.~Osman, S.~Kumar, A.~Sengar, A.~Chang, and A.~Alkhateeb, ``Design and evaluation of reconfigurable intelligent surfaces in real-world environment,'' \emph{IEEE Open J. Commun. Soc.}, vol.~3, no.~3, pp. 462--474, Mar. 2022.

\bibitem{TAP_2022_Liang_An}
J.~C. Liang, Q.~Cheng, Y.~Gao, C.~Xiao, S.~Gao, L.~Zhang, S.~Jin, and T.~J. Cui, ``An angle-insensitive 3-bit reconfigurable intelligent surface,'' \emph{IEEE Trans. Antennas Propag.}, vol.~70, no.~10, pp. 8798--8808, Oct. 2022.

\bibitem{OJCOMS_2024_Zhao_2}
Y.~Zhao, Y.~Feng, A.~M. Ismail, Z.~Wang, Y.~L. Guan, Y.~Guo, and C.~Yuen, ``2-bit {RIS} prototyping enhancing rapid-response space-time wavefront manipulation for wireless communication: Experimental studies,'' \emph{IEEE Open J. Commun. Society}, vol.~5, no.~1, pp. 4885--4901, Aug. 2024.

\bibitem{TAP_2024_Yang_High}
H.~Yang, Q.~Hu, Y.~Rao, W.~Zhong, and X.~Y. Zhang, ``High-flexibility low-complexity reprogrammable phase-continuous metasurface,'' \emph{IEEE Trans. Antennas Propag.}, vol.~72, no.~9, pp. 7146--7153, Sep. 2024.

\bibitem{TCOM_2021_Najafi_Physics}
M.~Najafi, V.~Jamali, R.~Schober, and H.~V. Poor, ``Physics-based modeling and scalable optimization of large intelligent reflecting surfaces,'' \emph{IEEE Trans. Commun.}, vol.~69, no.~4, pp. 2673--2691, Apr. 2021.

\bibitem{WCL_2021_Bjornson_Rayleigh}
E.~Björnson and L.~Sanguinetti, ``Rayleigh fading modeling and channel hardening for reconfigurable intelligent surfaces,'' \emph{IEEE Wireless Commun. Lett.}, vol.~10, no.~4, pp. 830--834, Apr. 2021.

\bibitem{Proc_2022_Renzo_Communication}
M.~Di~Renzo, F.~H. Danufane, and S.~Tretyakov, ``Communication models for reconfigurable intelligent surfaces: From surface electromagnetics to wireless networks optimization,'' \emph{Proc. IEEE}, vol. 110, no.~9, pp. 1164--1209, Sep. 2022.

\bibitem{TCOM_2022_Tang_Path}
W.~Tang, X.~Chen, M.~Z. Chen, J.~Y. Dai, Y.~Han, M.~D. Renzo, S.~Jin, Q.~Cheng, and T.~J. Cui, ``Path loss modeling and measurements for reconfigurable intelligent surfaces in the millimeter-wave frequency band,'' \emph{IEEE Trans. Commun.}, vol.~70, no.~9, pp. 6259--6276, Sep. 2022.

\bibitem{TWC_2022_Shen_Modeling}
S.~Shen, B.~Clerckx, and R.~Murch, ``Modeling and architecture design of reconfigurable intelligent surfaces using scattering parameter network analysis,'' \emph{IEEE Trans. Wireless Commun.}, vol.~21, no.~2, pp. 1229--1243, Feb. 2022.

\bibitem{TVT_2024_An_Adjustable}
J.~An, C.~Xu, D.~W.~K. Ng, C.~Yuen, and L.~Hanzo, ``Adjustable-delay {RIS} is capable of improving {OFDM} systems,'' \emph{IEEE Trans. Veh. Technol.}, vol.~73, no.~7, pp. 9927--9942, Jul. 2024.

\bibitem{JSAC_2020_Yu_Robust}
X.~Yu, D.~Xu, Y.~Sun, D.~W.~K. Ng, and R.~Schober, ``Robust and secure wireless communications via intelligent reflecting surfaces,'' \emph{IEEE J. Sel. Areas Commun.}, vol.~38, no.~11, pp. 2637--2652, Nov. 2020.

\bibitem{JSAC_2020_Bai_Latency}
T.~Bai, C.~Pan, Y.~Deng, M.~Elkashlan, A.~Nallanathan, and L.~Hanzo, ``Latency minimization for intelligent reflecting surface aided mobile edge computing,'' \emph{IEEE J. Sel. Areas Commun.}, vol.~38, no.~11, pp. 2666--2682, Nov. 2020.

\bibitem{JSAC_2020_Pizzo_Spatially}
A.~Pizzo, T.~L. Marzetta, and L.~Sanguinetti, ``Spatially-stationary model for holographic {MIMO} small-scale fading,'' \emph{IEEE J. Sel. Areas Commun.}, vol.~38, no.~9, pp. 1964--1979, Sep. 2020.

\bibitem{TSP_2018_Hu_Beyond}
S.~Hu, F.~Rusek, and O.~Edfors, ``Beyond massive {MIMO}: The potential of data transmission with large intelligent surfaces,'' \emph{IEEE Trans. Signal Process.}, vol.~66, no.~10, pp. 2746--2758, May 2018.

\bibitem{WC_2024_An_Near}
J.~An, C.~Yuen, L.~Dai, M.~Di~Renzo, M.~Debbah, and L.~Hanzo, ``Near-field communications: Research advances, potential, and challenges,'' \emph{IEEE Wireless Commun.}, vol.~31, no.~3, pp. 100--107, Jun. 2024.

\bibitem{TWC_2021_Wong_Fluid}
K.-K. Wong, A.~Shojaeifard, K.-F. Tong, and Y.~Zhang, ``Fluid antenna systems,'' \emph{IEEE Trans. Wireless Commun.}, vol.~20, no.~3, pp. 1950--1962, Mar. 2021.

\bibitem{TWC_2024_Zhu_Modeling}
L.~Zhu, W.~Ma, and R.~Zhang, ``Modeling and performance analysis for movable antenna enabled wireless communications,'' \emph{IEEE Trans. Wireless Commun.}, vol.~23, no.~6, pp. 6234--6250, Jun. 2024.

\bibitem{CL_2023_An_A}
J.~An, C.~Yuen, C.~Huang, M.~Debbah, H.~Vincent~Poor, and L.~Hanzo, ``A tutorial on holographic {MIMO} communications—{Part I}: Channel modeling and channel estimation,'' \emph{IEEE Commun. Lett.}, vol.~27, no.~7, pp. 1664--1668, Jul. 2023.

\bibitem{TWC_2020_Jung_Performance}
M.~Jung, W.~Saad, Y.~Jang, G.~Kong, and S.~Choi, ``Performance analysis of large intelligent surfaces ({LISs}): Asymptotic data rate and channel hardening effects,'' \emph{IEEE Trans. Wireless Commun.}, vol.~19, no.~3, pp. 2052--2065, Mar. 2020.

\bibitem{JSAC_2022_Deng_HDMA}
R.~Deng, B.~Di, H.~Zhang, and L.~Song, ``{HDMA}: Holographic-pattern division multiple access,'' \emph{IEEE J. Sel. Areas Commun.}, vol.~40, no.~4, pp. 1317--1332, Apr. 2022.

\bibitem{TWC_2023_Wei_Tri}
L.~Wei, C.~Huang, G.~C. Alexandropoulos, Z.~Yang, J.~Yang, W.~E.~I. Sha, Z.~Zhang, M.~Debbah, and C.~Yuen, ``Tri-polarized holographic {MIMO} surfaces for near-field communications: Channel modeling and precoding design,'' \emph{IEEE Trans. Wireless Commun.}, vol.~22, no.~12, pp. 8828--8842, Dec. 2023.

\bibitem{arXiv_2024_Dardari_Dynamic}
D.~Dardari, ``Dynamic scattering arrays for simultaneous electromagnetic processing and radiation in holographic {MIMO} systems,'' \emph{arXiv preprint arXiv:2405.16174}, 2024.

\bibitem{CST_2024_Gong_Holographic}
T.~Gong, P.~Gavriilidis, R.~Ji, C.~Huang, G.~C. Alexandropoulos, L.~Wei, Z.~Zhang, M.~Debbah, H.~V. Poor, and C.~Yuen, ``Holographic {MIMO} communications: Theoretical foundations, enabling technologies, and future directions,'' \emph{IEEE Commun. Surveys Tuts.}, vol.~26, no.~1, pp. 196--257, 1st Quarter 2024.

\bibitem{TWC_2021_Long_Active}
R.~Long, Y.-C. Liang, Y.~Pei, and E.~G. Larsson, ``Active reconfigurable intelligent surface-aided wireless communications,'' \emph{IEEE Trans. Wireless Commun.}, vol.~20, no.~8, pp. 4962--4975, Aug. 2021.

\bibitem{TCOM_2023_Zhang_Active}
Z.~Zhang, L.~Dai, X.~Chen, C.~Liu, F.~Yang, R.~Schober, and H.~V. Poor, ``Active {RIS} vs. passive {RIS}: Which will prevail in {6G}?'' \emph{IEEE Trans. Commun.}, vol.~71, no.~3, pp. 1707--1725, Mar. 2023.

\bibitem{TWC_2022_Mu_Simultaneously}
X.~Mu, Y.~Liu, L.~Guo, J.~Lin, and R.~Schober, ``Simultaneously transmitting and reflecting ({STAR}) {RIS} aided wireless communications,'' \emph{IEEE Trans. Wireless Commun.}, vol.~21, no.~5, pp. 3083--3098, May 2022.

\bibitem{CM_2022_Zhang_Intelligent}
H.~Zhang, S.~Zeng, B.~Di, Y.~Tan, M.~Di~Renzo, M.~Debbah, Z.~Han, H.~V. Poor, and L.~Song, ``Intelligent omni-surfaces for full-dimensional wireless communications: Principles, technology, and implementation,'' \emph{IEEE Commun. Mag.}, vol.~60, no.~2, pp. 39--45, Feb. 2022.

\bibitem{TWC_2022_Zhang_Intelligent}
S.~Zhang, H.~Zhang, B.~Di, Y.~Tan, M.~Di~Renzo, Z.~Han, H.~Vincent~Poor, and L.~Song, ``Intelligent omni-surfaces: Ubiquitous wireless transmission by reflective-refractive metasurfaces,'' \emph{IEEE Trans. Wireless Commun.}, vol.~21, no.~1, pp. 219--233, Jan. 2022.

\bibitem{JSAC_2020_You_Channel}
C.~You, B.~Zheng, and R.~Zhang, ``Channel estimation and passive beamforming for intelligent reflecting surface: Discrete phase shift and progressive refinement,'' \emph{IEEE J. Sel. Areas Commun.}, vol.~38, no.~11, pp. 2604--2620, Nov. 2020.

\bibitem{TIT_2023_Zhi_Two}
K.~Zhi, C.~Pan, H.~Ren, K.~Wang, M.~Elkashlan, M.~D. Renzo, R.~Schober, H.~V. Poor, J.~Wang, and L.~Hanzo, ``Two-timescale design for reconfigurable intelligent surface-aided massive {MIMO} systems with imperfect {CSI},'' \emph{IEEE Trans. Inf. Theory}, vol.~69, no.~5, pp. 3001--3033, May 2023.

\bibitem{TVT_2023_Xu_Channel}
C.~Xu, J.~An, T.~Bai, S.~Sugiura, R.~G. Maunder, Z.~Wang, L.-L. Yang, and L.~Hanzo, ``Channel estimation for reconfigurable intelligent surface assisted high-mobility wireless systems,'' \emph{IEEE Trans. Veh. Technol.}, vol.~72, no.~1, pp. 718--734, Jan. 2023.

\bibitem{TCCN_2022_Xu_Time}
W.~Xu, J.~An, Y.~Xu, C.~Huang, L.~Gan, and C.~Yuen, ``Time-varying channel prediction for {RIS}-assisted {MU-MISO} networks via deep learning,'' \emph{IEEE Trans. Cognitive Commun. Networking}, vol.~8, no.~4, pp. 1802--1815, Apr. 2022.

\bibitem{TSP_2020_Zhou_A}
G.~Zhou, C.~Pan, H.~Ren, K.~Wang, and A.~Nallanathan, ``A framework of robust transmission design for {IRS}-aided {MISO} communications with imperfect cascaded channels,'' \emph{IEEE Trans. Signal Process.}, vol.~68, pp. 5092--5106, Aug. 2020.

\bibitem{TGCN_2022_An_Joint}
J.~An, C.~Xu, L.~Wang, Y.~Liu, L.~Gan, and L.~Hanzo, ``Joint training of the superimposed direct and reflected links in reconfigurable intelligent surface assisted multiuser communications,'' \emph{IEEE Trans. Green Commun. Netw.}, vol.~6, no.~2, pp. 739--754, Feb. 2022.

\bibitem{TWC_2023_Ren_Configuring}
S.~Ren, K.~Shen, Y.~Zhang, X.~Li, X.~Chen, and Z.-Q. Luo, ``Configuring intelligent reflecting surface with performance guarantees: Blind beamforming,'' \emph{IEEE Trans. Wireless Commun.}, vol.~22, no.~5, pp. 3355--3370, May 2023.

\bibitem{TCOM_2024_Yu_Environment}
Z.~Yu, J.~An, E.~Basar, L.~Gan, and C.~Yuen, ``Environment-aware codebook design for {RIS}-assisted {MU-MISO} communications: Implementation and performance analysis,'' \emph{IEEE Trans. Commun.}, pp. 1--15, 2024, Early Access.

\bibitem{TAP_2024_Jia_High}
Y.~Jia, H.~Lu, Z.~Fan, B.~Wu, F.~Qu, M.-J. Zhao, C.~Qian, and H.~Chen, ``High-efficiency transmissive tunable metasurfaces for binary cascaded diffractive layers,'' \emph{IEEE Trans. Antennas Propag.}, vol.~72, no.~5, pp. 4532--4540, May 2024.

\bibitem{LSA_2020_Qian_Performing}
C.~Qian, X.~Lin, X.~Lin, J.~Xu, Y.~Sun, E.~Li, B.~Zhang, and H.~Chen, ``Performing optical logic operations by a diffractive neural network,'' \emph{Light: Science \& Applications}, vol.~9, no.~1, p.~59, Apr. 2020.

\bibitem{LSA_2024_Gao_Super}
S.~Gao, H.~Chen, Y.~Wang, Z.~Duan, H.~Zhang, Z.~Sun, Y.~Shen, and X.~Lin, ``Super-resolution diffractive neural network for all-optical direction of arrival estimation beyond diffraction limits,'' \emph{Light: Science \& Applications}, vol.~13, no.~1, p. 161, Jul. 2024.

\bibitem{arXiv_2024_Wang_Multi}
Z.~Wang, H.~Liu, J.~Zhang, R.~Xiong, K.~Wan, X.~Qian, M.~Di~Renzo, and R.~C. Qiu, ``Multi-user {ISAC} through stacked intelligent metasurfaces: New algorithms and experiments,'' \emph{arXiv preprint arXiv:2405.01104}, 2024.

\bibitem{TAP_2022_Liu_Prior}
P.~Liu, L.~Chen, and Z.~N. Chen, ``Prior-knowledge-guided deep-learning-enabled synthesis for broadband and large phase shift range metacells in metalens antenna,'' \emph{IEEE Trans. Antennas Propag.}, vol.~70, no.~7, pp. 5024--5034, Jul. 2022.

\bibitem{TAP_2023_Liu_Full}
P.~Liu and Z.~N. Chen, ``Full-range amplitude–phase metacells for sidelobe suppression of metalens antenna using prior-knowledge-guided deep-learning-enabled synthesis,'' \emph{IEEE Trans. Antennas Propag.}, vol.~71, no.~6, pp. 5036--5045, Jun. 2023.

\bibitem{Nano_2024_Lv_Meta}
Q.~Lv, X.~Qin, M.~Hu, P.~Li, Y.~Zhang, and Y.~Li, ``Metatronics-inspired high-selectivity metasurface filter,'' \emph{Nanophotonics}, no.~0, Apr. 2024.

\bibitem{ICC_2023_An_Stacked}
J.~An, M.~Di~Renzo, M.~Debbah, and C.~Yuen, ``Stacked intelligent metasurfaces for multiuser beamforming in the wave domain,'' in \emph{Proc. IEEE Int. Conf. Commun. (ICC)}, May 2023, pp. 2834--2839.

\bibitem{arXiv_2023_An_Stacked}
J.~An, M.~D. Renzo, M.~Debbah, H.~Vincent~Poor, and C.~Yuen, ``Stacked intelligent metasurfaces for multiuser downlink beamforming in the wave domain,'' \emph{IEEE Trans. Wireless Commun.}, pp. 1--15, 2025, Early Access.

\bibitem{arXiv_2024_Darsena_Design}
D.~Darsena, F.~Verde, I.~Iudice, and V.~Galdi, ``Design of stacked intelligent metasurfaces with reconfigurable amplitude and phase for multiuser downlink beamforming,'' \emph{arXiv preprint arXiv:2408.16606}, 2024.

\bibitem{WCL_2024_Papazafeiropoulos_Achievable}
A.~Papazafeiropoulos, P.~Kourtessis, S.~Chatzinotas, D.~I. Kaklamani, and I.~S. Venieris, ``Achievable rate optimization for large stacked intelligent metasurfaces based on statistical {CSI},'' \emph{IEEE Wireless Commun. Lett.}, vol.~13, no.~9, pp. 2337--2341, Sep. 2024.

\bibitem{arXiv_2024_Rezvani_Uplink}
M.~Rezvani, R.~Adve, A.~b. Sediq, and A.~El-Keyi, ``Uplink wave-domain combiner for stacked intelligent metasurfaces accounting for hardware limitations,'' \emph{arXiv preprint arXiv:2407.21012}, 2024.

\bibitem{CL_2024_Nerini_Physically}
M.~Nerini and B.~Clerckx, ``Physically consistent modeling of stacked intelligent metasurfaces implemented with beyond diagonal {RIS},'' \emph{IEEE Commun. Lett.}, vol.~28, no.~7, pp. 1693--1697, Jul. 2024.

\bibitem{arXiv_2024_Papazafeiropoulos_Performance}
A.~Papazafeiropoulos, P.~Kourtessis, S.~Chatzinotas, D.~I. Kaklamani, and I.~S. Venieris, ``Performance of double-stacked intelligent metasurface-assisted multiuser massive {MIMO} communications in the wave domain,'' \emph{IEEE Trans. Wireless Commun.}, pp. 1--13, 2025, Early Access.

\bibitem{APWCS_2024_Li_Stacked}
Z.~Li, J.~An, and C.~Yuen, ``Stacked intelligent metasurfaces for fully-analog wideband beamforming design,'' in \emph{Proc. IEEE VTS Asia Pacific Wireless Commun. Sym. (APWCS)}, Aug. 2024, pp. 1--5.

\bibitem{TWC_2024_Papazafeiropoulos_Achievable}
A.~Papazafeiropoulos, J.~An, P.~Kourtessis, T.~Ratnarajah, and S.~Chatzinotas, ``Achievable rate optimization for stacked intelligent metasurface-assisted holographic {MIMO} communications,'' \emph{IEEE Trans. Wireless Commun.}, vol.~23, no.~10, pp. 13\,173--13\,186, Oct. 2024.

\bibitem{CL_2024_Perovic_Mutual}
N.~S. Perović and L.-N. Tran, ``Mutual information optimization for {SIM}-based holographic {MIMO} systems,'' \emph{IEEE Commun. Lett.}, pp. 1--5, 2024, Early Access.

\bibitem{arXiv_2024_Perovic_Energy}
N.~S. Perovi{\'c}, E.~E. Bahingayi, and L.-N. Tran, ``Energy-efficient designs for {SIM}-based broadcast {MIMO} systems,'' \emph{arXiv preprint arXiv:2409.00628}, 2024.

\bibitem{WCL_2024_Lin_Stacked}
S.~Lin, J.~An, L.~Gan, M.~Debbah, and C.~Yuen, ``Stacked intelligent metasurface enabled {LEO} satellite communications relying on statistical {CSI},'' \emph{IEEE Wireless Commun. Lett.}, vol.~13, no.~5, pp. 1295--1299, May 2024.

\bibitem{arXiv_2024_Liu_Multi}
H.~Liu, J.~An, G.~C. Alexandropoulos, D.~W.~K. Ng, C.~Yuen, and L.~Gan, ``Multi-user {MISO} with stacked intelligent metasurfaces: A {DRL}-based sum-rate optimization approach,'' \emph{IEEE Trans. Cognitive Commun. Netw.}, pp. 1--14, 2025, Early Access.

\bibitem{arXiv_2024_Yang_Joint}
X.~Yang, J.~Zhang, E.~Shi, Z.~Liu, J.~Liu, K.~Zheng, and B.~Ai, ``Joint {SIM} configuration and power allocation for stacked intelligent metasurface-assisted {MU-MISO} systems with {TD3},'' \emph{arXiv preprint arXiv:2408.05756}, 2024.

\bibitem{arXiv_2024_Amirhosein_Meta}
A.~Mohammadzadeh, H.~Zarini, M.~Robat~Mili, M.~Jafari~Siavoshani, A.~Movaghar, J.~An, and N.~Al-Dhahir, ``Meta reinforcement learning empowered orchestration of {SIM} and {RIS} for downlink multiuser communications,'' \emph{arXiv preprint}, 2024.

\bibitem{WCL_2024_Papazafeiropoulos_Near}
A.~Papazafeiropoulos, P.~Kourtessis, S.~Chatzinotas, D.~I. Kaklamani, and I.~S. Venieris, ``Near-field beamforming for stacked intelligent metasurfaces-assisted {MIMO} networks,'' \emph{IEEE Wireless Commun. Lett.}, pp. 1--5, 2024, Early Access.

\bibitem{VTC_2024_Jia_Stacked}
X.~Jia, J.~An, H.~Liu, L.~Gan, M.~Di~Renzo, M.~Debbah, and C.~Yuen, ``Stacked intelligent metasurface enabled near-field multiuser beamfocusing in the wave domain,'' in \emph{Proc. IEEE 99th Veh. Technol. Conf. (VTC2024-Spring)}, Jun. 2024, pp. 1--5.

\bibitem{APWCS_2024_Yao_Sparse}
X.~Yao, J.~An, G.~Huang, H.~Liu, L.~Gan, and C.~Yuen, ``Sparse channel estimation for stacked intelligent metasurface-assisted mmwave communications,'' in \emph{Proc. IEEE VTS Asia Pacific Wireless Commun. Sym. (APWCS)}, Aug. 2024, pp. 1--5.

\bibitem{WCL_2024_Yao_Channel}
X.~Yao, J.~An, L.~Gan, M.~Di~Renzo, and C.~Yuen, ``Channel estimation for stacked intelligent metasurface-assisted wireless networks,'' \emph{IEEE Wireless Commun. Lett.}, vol.~13, no.~5, pp. 1349--1353, May 2024.

\bibitem{WCNC_2024_Nadeem_Hybrid}
Q.-U.-A. Nadeem, J.~An, and A.~Chaaban, ``Hybrid digital-wave domain channel estimator for stacked intelligent metasurface enabled multi-user {MISO} systems,'' in \emph{2024 IEEE Wireless Communications and Networking Conference (WCNC)}, Apr. 2024, pp. 1--6.

\bibitem{TCOM_2024_Li_Stacked}
Q.~Li, M.~El-Hajjar, C.~Xu, J.~An, C.~Yuen, and L.~Hanzo, ``Stacked intelligent metasurfaces for holographic {MIMO} aided cell-free networks,'' \emph{IEEE Trans. Commun.}, pp. 1--15, 2024, Early Access.

\bibitem{arXiv_2024_Shi_Harnessing}
E.~Shi, J.~Zhang, Y.~Zhu, J.~An, C.~Yuen, and B.~Ai, ``Uplink performance of stacked intelligent metasurface-enhanced cell-free massive {MIMO} systems,'' \emph{IEEE Trans. Wireless Commun.}, pp. 1--15, 2025, Early Access.

\bibitem{arXiv_2024_Shi_Joint}
E.~Shi, J.~Zhang, J.~An, G.~Zhang, Z.~Liu, C.~Yuen, and B.~Ai, ``Joint {AP-UE} association and precoding for {SIM}-aided cell-free massive {MIMO} systems,'' \emph{IEEE Trans. Wireless Commun.}, pp. 1--15, 2025, Early Access.

\bibitem{WCL_2024_Niu_Stacked}
H.~Niu, J.~An, A.~Papazafeiropoulos, L.~Gan, S.~Chatzinotas, and M.~Debbah, ``Stacked intelligent metasurfaces for integrated sensing and communications,'' \emph{IEEE Wireless Commun. Lett.}, vol.~13, no.~10, pp. 2807--2811, Oct. 2024.

\bibitem{arXiv_2024_Li_Transmit}
S.~Li, F.~Zhang, T.~Mao, R.~Na, Z.~Wang, and G.~K. Karagiannidis, ``Transmit beamforming design for {ISAC} with stacked intelligent metasurfaces,'' \emph{IEEE Trans. Veh. Technol.}, vol.~74, no.~4, pp. 6767--6772, 2025.

\bibitem{arXiv_2024_Huang_Stacked}
G.~Huang, J.~An, Z.~Yang, L.~Gan, M.~Bennis, and M.~Debbah, ``Stacked intelligent metasurfaces for task-oriented semantic communications,'' \emph{IEEE Wireless Commun. Lett.}, vol.~14, no.~2, pp. 310--314, 2025.

\bibitem{OJCOMS_2024_Hassan_Efficient}
N.~U. Hassan, J.~An, M.~Di~Renzo, M.~Debbah, and C.~Yuen, ``Efficient beamforming and radiation pattern control using stacked intelligent metasurfaces,'' \emph{IEEE Open J. Commun. Soc.}, vol.~5, pp. 599--611, May 2024.

\bibitem{APWCS_2024_Niu_Enhancing}
H.~Niu, J.~An, L.~Zhang, X.~Lei, and C.~Yuen, ``Enhancing physical layer security for {SISO} systems using stacked intelligent metasurfaces,'' in \emph{Proc. IEEE VTS Asia Pacific Wireless Commun. Sym. (APWCS)}, Aug. 2024, pp. 1--5.

\bibitem{ICC_2024_Liu_DRL}
H.~Liu, J.~An, D.~W.~K. Ng, G.~C. Alexandropoulos, and L.~Gan, ``{DRL}-based orchestration of multi-user {MISO} systems with stacked intelligent metasurfaces,'' in \emph{Proc. IEEE Int. Conf. Commun.}, Jun. 2024, pp. 4991--4996.

\bibitem{IoTJ_2024_Xu_Algorithm}
W.~Xu, J.~An, H.~Li, L.~Gan, and C.~Yuen, ``Algorithm-unrolling-based distributed optimization for {RIS}-assisted cell-free networks,'' \emph{IEEE Int. Things J.}, vol.~11, no.~1, pp. 944--957, Jan. 2024.

\bibitem{Proc_2024_Shi_RIS}
E.~Shi, J.~Zhang, H.~Du, B.~Ai, C.~Yuen, D.~Niyato, K.~B. Letaief, and X.~Shen, ``{RIS}-aided cell-free massive {MIMO} systems for {6G}: Fundamentals, system design, and applications,'' \emph{Proc. IEEE}, vol. 112, no.~4, pp. 331--364, Apr. 2024.

\bibitem{APWCS_2024_Shi_Uplink}
E.~Shi, J.~Zhang, Y.~Zhu, Z.~Liu, J.~An, C.~Yuen, and B.~Ai, ``Uplink performance and beamforming design of {SIM}-enhanced cell-free massive {MIMO} systems,'' in \emph{Proc. IEEE VTS Asia Pacific Wireless Commun. Symposium (APWCS)}, Aug. 2024, pp. 01--05.

\bibitem{IoTJ_2023_Xu_OTFS}
C.~Xu, L.~Xiang, J.~An, C.~Dong, S.~Sugiura, R.~G. Maunder, L.-L. Yang, and L.~Hanzo, ``{OTFS}-aided {RIS}-assisted {SAGIN} systems outperform their {OFDM} counterparts in doubly selective high-doppler scenarios,'' \emph{IEEE Int. Things J.}, vol.~10, no.~1, pp. 682--703, Jan. 2023.

\bibitem{TWC_2023_An_Fundamental}
J.~An, H.~Li, D.~W.~K. Ng, and C.~Yuen, ``Fundamental detection probability vs. achievable rate tradeoff in integrated sensing and communication systems,'' \emph{IEEE Trans. Wireless Commun.}, vol.~22, no.~12, pp. 9835--9853, Dec. 2023.

\bibitem{ICC_2024_An_Stacked}
J.~An, C.~Yuen, Y.~L. Guan, M.~Di~Renzo, M.~Debbah, H.~V. Poor, and L.~Hanzo, ``Stacked intelligent metasurface performs a {2D DFT} in the wave domain for {DOA} estimation,'' in \emph{Proc. IEEE Int. Conf. Commun.}, Jun. 2024, pp. 3445--3451.

\bibitem{IJCNN_2017_Cohen_EMNIST}
G.~Cohen, S.~Afshar, J.~Tapson, and A.~Van~Schaik, ``{EMNIST}: Extending {MNIST} to handwritten letters,'' in \emph{Proc. Int. Joint Conf. Neural Netw. (IJCNN)}.\hskip 1em plus 0.5em minus 0.4em\relax IEEE, 2017, pp. 2921--2926.

\bibitem{JSTQE_2020_Mengu_Analysis}
D.~Mengu, Y.~Luo, Y.~Rivenson, and A.~Ozcan, ``Analysis of diffractive optical neural networks and their integration with electronic neural networks,'' \emph{IEEE J. Sel. Topics Quantum Elect.}, vol.~26, no.~1, pp. 1--14, Jan. 2020.

\bibitem{NE_2023_Gao_Programmable}
X.~Gao, Q.~Ma, Z.~Gu, W.~Y. Cui, C.~Liu, J.~Zhang, and T.~J. Cui, ``Programmable surface plasmonic neural networks for microwave detection and processing,'' \emph{Nature Electronics}, vol.~6, no.~4, pp. 319--328, Apr. 2023.

\bibitem{NP_2024_Yildirim_Nonlinear}
M.~Yildirim, N.~U. Dinc, I.~Oguz, D.~Psaltis, and C.~Moser, ``Nonlinear processing with linear optics,'' \emph{Nature Photonics}, vol.~18, no.~10, pp. 1076--1082, Jul. 2024.

\bibitem{APS_2021_Luo_Electromagnetic}
X.~Luo, M.~Pu, Y.~Guo, X.~Li, and X.~Ma, ``Electromagnetic architectures: Structures, properties, functions and their intrinsic relationships in subwavelength optics and electromagnetics,'' \emph{Advanced Photonics Research}, vol.~2, no.~10, p. 2100023, May 2021.

\bibitem{TAP_2022_Benoni_Planning}
A.~Benoni, M.~Salucci, G.~Oliveri, P.~Rocca, B.~Li, and A.~Massa, ``Planning of {EM} skins for improved quality-of-service in urban areas,'' \emph{IEEE Trans. Antennas Propag.}, vol.~70, no.~10, pp. 8849--8862, Oct. 2022.

\bibitem{ACS_2017_Cong_Perovskite}
L.~Cong, Y.~K. Srivastava, A.~Solanki, T.~C. Sum, and R.~Singh, ``Perovskite as a platform for active flexible metaphotonic devices,'' \emph{ACS Photo.}, vol.~4, no.~7, pp. 1595--1601, May 2017.

\bibitem{NP_2020_Qian_Deep}
C.~Qian, B.~Zheng, Y.~Shen, L.~Jing, E.~Li, L.~Shen, and H.~Chen, ``Deep-learning-enabled self-adaptive microwave cloak without human intervention,'' \emph{Nature Photo.}, vol.~14, no.~6, pp. 383--390, Mar. 2020.

\bibitem{Small_2023_Nam_Flexible}
J.~Nam, I.~Chang, J.-S. Lim, H.~Woo, J.-G. Yook, and H.~H. Cho, ``Flexible metasurface for microwave-infrared compatible camouflage via particle swarm optimization algorithm,'' \emph{Small}, vol.~19, no.~2, pp. 1--9, Jun. 2023.

\bibitem{SR_2016_Zhang_Broadband}
Y.~Zhang, L.~Liang, J.~Yang, Y.~Feng, B.~Zhu, J.~Zhao, T.~Jiang, B.~Jin, and W.~Liu, ``Broadband diffuse terahertz wave scattering by flexible metasurface with randomized phase distribution,'' \emph{Scientific Reports}, vol.~6, no.~1, p. 26875, May 2016.

\bibitem{OE_2016_Zhao_Achieving}
J.~Zhao, B.~Sima, N.~Jia, C.~Wang, B.~Zhu, T.~Jiang, and Y.~Feng, ``Achieving flexible low-scattering metasurface based on randomly distribution of meta-elements,'' \emph{Optics Express}, vol.~24, no.~24, pp. 27\,849--27\,857, Nov. 2016.

\bibitem{AOM_2017_Song_Water}
Q.~Song, W.~Zhang, P.~C. Wu, W.~Zhu, Z.~X. Shen, P.~H.~J. Chong, Q.~X. Liang, Z.~C. Yang, Y.~L. Hao, H.~Cai \emph{et~al.}, ``Water-resonator-based metasurface: an ultrabroadband and near-unity absorption,'' \emph{Advanced Optical Materials}, vol.~5, no.~8, p. 1601103, Mar. 2017.

\bibitem{AM_2021_Wang_Pangolin}
C.~Wang, Z.~Lv, M.~P. Mohan, Z.~Cui, Z.~Liu, Y.~Jiang, J.~Li, C.~Wang, S.~Pan, M.~F. Karim \emph{et~al.}, ``Pangolin-inspired stretchable, microwave-invisible metascale,'' \emph{Adv. Materials}, vol.~33, no.~41, p. 2102131, Aug. 2021.

\bibitem{NC_2016_Kamali_Decoupling}
S.~M. Kamali, A.~Arbabi, E.~Arbabi, Y.~Horie, and A.~Faraon, ``Decoupling optical function and geometrical form using conformal flexible dielectric metasurfaces,'' \emph{Nature Communications}, vol.~7, no.~1, p. 11618, May 2016.

\bibitem{AM_2011_Aksu_Flexible}
S.~Aksu, M.~Huang, A.~Artar, A.~A. Yanik, S.~Selvarasah, M.~R. Dokmeci, and H.~Altug, ``Flexible plasmonics on unconventional and nonplanar substrates,'' \emph{Advanced Materials}, vol.~23, no.~38, p. 4422, Oct. 2011.

\bibitem{AS_2019_Huang_Catenary}
Y.~Huang, J.~Luo, M.~Pu, Y.~Guo, Z.~Zhao, X.~Ma, X.~Li, and X.~Luo, ``Catenary electromagnetics for ultra-broadband lightweight absorbers and large-scale flat antennas,'' \emph{Advanced Science}, vol.~6, no.~7, p. 1801691, Feb. 2019.

\bibitem{NC_2024_An_Energy}
S.~An, X.~Li, Z.~Guo, Y.~Huang, Y.~Zhang, and H.~Jiang, ``Energy-efficient dynamic {3D} metasurfaces via spatiotemporal jamming interleaved assemblies for tactile interfaces,'' \emph{Nature Communications}, vol.~15, no.~1, p. 7340, Aug. 2024.

\bibitem{NC_2019_Hajiesmaili_Reconfigurable}
E.~Hajiesmaili and D.~R. Clarke, ``Reconfigurable shape-morphing dielectric elastomers using spatially varying electric fields,'' \emph{Nature Communications}, vol.~10, no.~1, p. 183, Jan. 2019.

\bibitem{NC_2022_Ni_Soft}
X.~Ni, H.~Luan, J.-T. Kim, S.~I. Rogge, Y.~Bai, J.~W. Kwak, S.~Liu, D.~S. Yang, S.~Li, S.~Li \emph{et~al.}, ``Soft shape-programmable surfaces by fast electromagnetic actuation of liquid metal networks,'' \emph{Nature Communications}, vol.~13, no.~1, p. 5576, Sep. 2022.

\bibitem{SA_2023_Wang_Passively}
J.~Wang, M.~Sotzing, M.~Lee, and A.~Chortos, ``Passively addressed robotic morphing surface ({PARMS}) based on machine learning,'' \emph{Science Advances}, vol.~9, no.~29, p. eadg8019, Jul. 2023.

\bibitem{GLOBECOM_2025_An_Downlink}
J.~An, C.~Yuen, M.~Di~Renzo, M.~Debbah, H.~V. Poor, and L.~Hanzo, ``Downlink multiuser communications relying on flexible intelligent metasurfaces,'' \emph{arXiv preprint}, 2024.

\bibitem{arXiv_2024_Yang_Flexible}
S.~Yang, J.~An, Y.~Xiu, W.~Lyu, B.~Ning, Z.~Zhang, M.~Debbah, and C.~Yuen, ``Flexible antenna arrays for wireless communications: Modeling and performance evaluation,'' \emph{arXiv preprint arXiv:2407.04944}, 2024.

\end{thebibliography}
\bibliographystyle{IEEEtran}

\end{document}